# From specific-source feature-based to common-source score-based likelihood-ratio systems: ranking the stars

Peter Vergeer , Forensic Statistics & Big Data Analysis, Netherlands Forensic Institute, The Hague, Netherlands. p.vergeer@nfi.nl

## Abstract

This paper studies expected performance and practical feasibility of the most commonly used classes of source-level likelihood-ratio (LR) systems when applied to a trace-reference comparison problem. The paper compares performance of these classes of LR systems (used to update prior odds) to each other and to the use of prior odds only, using strictly proper scoring rules as performance measures. It also explores practical feasibility of the classes of LR systems. The present analysis allows for a ranking of these classes of LR systems: from specific-source feature-based to common-source anchored or non-anchored score-based. A trade-off between performance and practical feasibility is observed, meaning that the best performing class of LR systems is the hardest to realise in practice, while the least performing class is the easiest to realise in practice. The other classes of LR systems are in between the two extremes. The one positive exception is a common-source feature-based LR system, with good performance and relatively low experimental demands. The paper also argues against the claim that some classes of LR systems should not be used, by showing that all systems have merit (when updating prior odds) over just using the prior odds (i.e. not using the LR system).

# 1. Introduction

A likelihood ratio is widely accepted as the appropriate way to communicate evidential strength in forensic science and in court cases, for example in most parts of Europe (*ENFSI Guideline for Evaluative Reporting in Forensic Science (STEOFRAE)*, 2015), Australia & New Zealand (Catoggio et al., 2019), and, albeit with criticism (Iyer and Lund, 2017), to some degree in the United States (Levey, 2019).

A likelihood ratio is defined as the ratio of two probabilities of observing some evidence, given each of two competing hypotheses. These probabilities may be further conditioned on background information and other evidence in the case. Likelihood ratios can be used to update the prior odds of a hypothesis pair by using Bayes' rule to obtain its posterior odds. This framework also conveniently provides a division of roles of the actors in court. The forensic expert provides a likelihood ratio for the evidence while the judge/jury provides prior knowledge on the hypotheses based on case circumstances and may update this knowledge with the likelihood ratio conclusion of the forensic expert.

A recent trend in forensic science is to automatically compute a likelihood ratio based on mathematically specified models for the probabilities of the evidence trained by relevant data (Morrison and Stoel, 2014), for example see Aitken et al., 2019, 2007; Aitken and Lucy, 2004; Aitken and Taroni, 2004; Alberink et al., 2014; Bolck et al., 2009; Brümmer and Doddington, 2013; Franco-Pedroso et al., 2016; Garton, 2020; Geoffrey et al., 2020; Hendricks et al., 2018; Hepler et al., 2012; Lindley, 1977; Morrison and Enzinger, 2018; Neumann et al., 2015, 2012, 2006; Ommen et al., 2017; Stoney et al., 2020; van Es et al., 2017; Vergeer et al., 2014; Zadora et al., 2013. In this paper, such automated procedures are referred to as LR systems: automated, computer-based methods aimed at evaluating evidential value in court cases and that give likelihood ratios as output.

## 1.1. Different classes of LR systems

Part of the current forensic scientific research on LR systems concentrates on the appropriateness of different choices when designing these systems. Design choices are generally made on two dimensions: which type of hypotheses to consider and what evidence. Regarding the hypotheses, so-called specific-source and common-source hypothesis pairs are distinguished (Dorp et al., 2020; Garton, 2020; Ommen et al., 2017, 2015; Ommen and Saunders, 2019, 2018). Regarding the evidence, a spectrum with two extremes may be considered. On the one extreme, the measurement outcomes (or features) may be considered directly, and its distribution is statistically modelled. On the other extreme, a transformation of the features to a one-dimensional quantity called a 'score' (a scalar value expressing similarity or difference between measurements on two objects) is considered as the evidence (Vergeer et al., 2020a) and distributions of the score are modelled. In between are methods that reduce the number of features to a lower-dimensional space for which statistical models are considered, see for example Aitken et al., 2019; Vergeer et al., 2014.

When a score is considered (for example, a Euclidian distance metric) it can be used to compare measurement $x$ from trace objects to measurement $y$ from reference objects. The probability (density) of the score under both hypotheses is calculated when computing a likelihood ratio. Within the score-based type there is a distinction between unanchored and anchored likelihood ratio systems (Alberink et al., 2014; Hepler et al., 2012); the measured features of the reference or trace objects are considered fixed (anchored) or not (non-anchored). When an anchored score-based LR system is used, the distribution of the score given the measured feature used for anchoring is calculated.

With the advent of deep neural networks in forensic science, score-based LRs have become more important. Neural networks do not use well thought-out statistical model families but rely on (many) layers of artificial neurons and in the end transform $x$ and $y$ to a one-dimensional discriminative quantity (thus: a score). A suitable transformation for this score to LR needs to be found, see for example Ferrer et al., 2021.

### 1.2. Criticisms on certain classes of LR systems

In recent years, forensic scientists and practitioners have voiced some serious criticisms regarding certain classes of LR systems, advocating that they should not be used or their use should be restricted. The results of this paper argue against these claims. These criticisms encompass the following:

i. Common-source LR systems are irrelevant to questions in court when a suspect (or reference source) is identified (Ommen and Saunders, 2018, Neumann and et al., 2020). One of the most important questions in court is whether a trace object is from a specific source or not, whereas the common-source scenario is framed to answer the question whether two trace objects have a common source or not (Ommen and Saunders, 2018). In the latter case, the source is unknown and treated as a random variable while for specific-source questions the reference should be treated as known and fixed. This implies different statistical modelling to answer a different question.

ii. Some classes of LR systems may unpredictably over or underestimate the weight of the evidence represented by measurements on a trace and a reference object (Neumann and Ausdemore, 2020). When feature-based LRs are used as benchmark values, score-based LRs have a different numeric value, usually less extreme. This leads to the result that for the same measurements, LR values are different depending on which LR system is used, and score-based LRs may be biased against certain suspects. Therefore, score-based LR systems are generally misleading, possible biased, and should not be used.

iii. Score-based LRs that do not include a measure of rarity do not qualify as appropriate likelihood ratios (Morrison and Enzinger, 2018). When information about typicality (i.e. rarity) of a trace and/or reference measurement is not captured in the score, the score merely expresses similarity between measurements on a trace and a reference object. Such a score is an incomplete summary of the evidence since, given the same score, common and rare references get the same likelihood ratio. This is wrong since a likelihood ratio should capture rarity as well (Morrison and Enzinger, 2018; Neumann and et al., 2020; Neumann and Ausdemore, 2020).

If these criticisms hold ground, this would disqualify a large number of LR systems featured in recent scientific literature: all LR systems that are either based on common-source hypotheses for the trace-reference problem (Aitken et al., 2019, 2007; Aitken and Lucy, 2004; Meester and Slooten, 2020; Neumann et al., 2006; Vergeer et al., 2020b, 2014) and/or score-based LR systems that do not include rarity (Bolck et al., 2009; Hepler et al., 2012; Leegwater et al., 2017; Vergeer et al., 2014), and/or LR systems that are based on neural networks (Ferrer et al., 2021; Rodriguez et al., 2020) for which it is unknown how (and maybe if) rarity is included.

### 1.3. Trigger for this study

The trigger for this study is the Law, Probability and Risk paper of Ommen and Saunders (Ommen and Saunders, 2018), where they give an overview of the difference between specific and common-source problems, hypotheses and sampling models. While reading their paper, I learned a few things, but other things leaved me confused.

I learned that:

1. Hypotheses do not consider the source of evidence, but the source of objects. And measurements from the objects are used as evidence for inference of source. Objects come from sources. For example: a bullet (object) is fired from a gun (source), or a glass fragment (object) is from a window (source).
2. The authors distinguish measurement models, which address probabilistic models for measurements given objects, from sampling models, which address sampling objects from a source or sampling sources from a population of sources. Also, they note that measurement models are generally not disputed, it is the sampling models that are disputed. A likelihood ratio distinguishes two competing mathematical distributional models based on the sampling models, not two competing hypotheses; the hypotheses are represented by the mathematical models.

In the sections that follow, I refer to a collection of objects from sources as an object collection. A measurement set is a set of measurements partitioned over the objects.

Ommen and Saunders consider the specific-source problem:

- We have an object collection from one trace source, we call this collection $o_t$.
- We have an object collection from one known reference source, we call this collection $o_r$.
- It is disputed whether $o_t$ is from the known reference source.

They contrast this with the common-source problem:

- We have an object collection from one trace source 1, we call this collection $o_{t1}$.
- We have an object collection from one trace source 2, we call this collection $o_{t2}$.
- It is disputed whether $o_{t1}$ and $o_{t2}$ have a common, unknown source, or whether they have different sources.

Using hypotheses on the object level, the hypotheses used by Ommen and Saunders (Ommen and Saunders, 2018) read:

Hypotheses: specific source

$H_1$: $o_r$ and $o_t$ both originate from the specific source.

$H_2$: $o_t$ does not originate from the specific source, but it originates from another source from the alternative source population. $o_r$ originates from the specific source.

Hypotheses: common source

$H_1$: $o_{t1}$ and $o_{t2}$ both originate from the same, unknown source.

$H_2$: $o_{t1}$ and $o_{t2}$ originate from two different unknown sources.

As Ommen and Saunders point out, the conditioning terms in the likelihood ratios should not only be distinguished by the hypotheses, but also by their corresponding sampling models.

So let's look at the specification of the sampling models for the specific and the common-source problem as put forward by Ommen and Saunders in (Ommen and Saunders, 2021, 2018).

Sampling models: specific source

$M_s$: $o_r$ is generated by random selection from the population of objects from the specific source.

$M_a$: $o_a$ (the 'alternative sources object collection') is generated by first randomly selecting $n_a$ sources from the population of alternative sources and then randomly selecting $n_i$ objects from within the i-th source, for $i = 1, \ldots, n_a$.

$M_1$: $o_t$ is generated by random selection from the population of objects from the specific source.

$M_2$: $o_t$ is generated by first randomly selecting a new source from the population of alternative sources and then randomly selecting objects from within that source.

Sampling models: common source

$M_a$: $o_a$ is generated by first randomly selecting $n_a$ sources from the population of alternative sources and then randomly selecting $n_i$ objects from within the i-th source, for $i = 1, \ldots, n_a$ (the same as for the specific-source $M_a$ sampling model).

$M_1$: $o_{t1}$ and $o_{t2}$ are generated by first randomly selecting a new source from the population of alternative sources and then randomly selecting objects from within that source.

$M_2$: $o_{t1}$ and $o_{t2}$ are generated by first randomly selecting two different sources from the population of alternative sources, then $o_{t1}$ is generated by randomly selecting objects from within the first source and $o_{t2}$ is generated by randomly selecting objects from within the second source.

When comparing the sampling models, it stands out that the specific-source sampling model for $o_r$ possesses one less level of randomness as compared to the sampling models for $o_t$, $o_{t1}$, $o_{t2}$, or $o_a$. All sources for these latter object collections are selected randomly from some background population of sources while the specific source stands disjointed; it is not selected from some background source population. This is not only the case in (Ommen and Saunders, 2018, 2021), but also in their more technical paper (Ommen et al., 2017), where a prior is used that directly describes the prior distribution of specific source characteristics (or: parameter values).

In the LPR paper (Ommen and Saunders, 2018), the authors exemplify the two different sampling models by using two simple, single-source, matching DNA profiles, for sources p1 and p2. Say the observed profile in common is $\Gamma$, with known frequency $\gamma$ in a relevant population.
The equation for the LR used is:

$$LR = \frac{P(\Gamma_{p2}|\Gamma_{p1}, H_1)}{P(\Gamma_{p2}|\Gamma_{p1}, H_2)} \times \frac{P(\Gamma_{p1}|H_1)}{P(\Gamma_{p1}|H_2)}$$

The term $\frac{P(\Gamma_{p2}|\Gamma_{p1}, H_1)}{P(\Gamma_{p2}|\Gamma_{p1}, H_2)}$ is not controversial and reads $\frac{1}{\gamma}$ for both common and specific-source LRs.

However, the mathematical expression for the right-hand term $\frac{P(\Gamma_{p1}|H_1)}{P(\Gamma_{p1}|H_2)}$ is different for common and specific-source approaches (although, as Ommen and Saunders note, the numerical outcome for the LR is identical). They reason that, under the specific-source hypotheses p1 is a known suspect (Mister X say). Therefore, it is inevitable that we observe Γ for p1 and its probability is 1 under both hypotheses. Then, the mathematical expression for the LR reads $\frac{1}{\gamma} \times \frac{1}{1}$.

Under the common-source hypotheses however, p1 is fundamentally unknown and selected randomly from a relevant background population and the known frequency of Γ is $\gamma$ in this population. Thus, the mathematical expression for the common-source LR reads $\frac{1}{\gamma} \times \frac{\gamma}{\gamma}$. An extra level of randomness is included in the right-hand term.

In (Ommen and Saunders, 2018) the authors conclude that common-source and specific-source problems are associated with different hypothesis pairs and sampling models, and likelihood ratios are generally different (although not in this DNA example, since the frequency of Γ in the relevant population of humans was assumed known and there is no randomness in the measurement model). Therefore, one should be careful not to confuse one with the other and the specific-source hypotheses are 'of most interest to a trial of justice' (Ommen and Saunders, 2018), since one usually has a known reference source. More explicitly, Neumann et al. (2020) conclude that in general, 'miss-specifying the interpretation framework can lead to dramatic results', since values for common-source and specific-source LRs are generally different when there is randomness in the measurement model.

Now this is what left me confused:

Having read this, I still wondered why, under the specific-source scenario (i.e. comparing a trace to a known source), the probability to observe a profile Γ for the known suspect must be 1. Ignoring measurement errors, it is inevitable that we observe this profile from the known suspect, which may have motivated the assignment of a probability of 1. However, from a Bayesian perspective the situation is different. After all, we may be able to identify the specific source as Mister X, but that does not mean we know his profile before measuring it. From a Bayesian perspective, the latent profile of Mister X is a nuisance parameter and needs to be integrated out.

So I wondered what would happen to the specific-source likelihood ratio when Mister X is identified, but the profile of Mister X is assumed unknown (before observing it by measurement). However, I wanted to avoid using subjective measures of a priori belief for parameters. An obvious assumption that meets this criterion is that the specific source (Mister X) is, as far as his latent DNA-profile is of concern, *a person selected at random from the relevant population of humans*. We also assume we know the frequency $\gamma$ of $\Gamma$ in that population (avoiding subjective measures of a priori belief). Now, the mathematical description of the specific-source LR is:

$$\frac{1}{\gamma} \times \frac{\gamma}{\gamma},$$

which is the same as for the common-source LR.

In doing so, we did identify Mister X as the specific source, but we have also:

a) added an hierarchical level using the distribution of $\Gamma$ between reference sources (it is random as far as we know, and its population frequency is $\gamma$). This means we use for the specific-source problem the common-source sampling model for $o_{t1}$.
b) this extra level of randomness results in a probability to observe the profile $\Gamma$ for Mister X of $\gamma$, and not 1.
c) the resulting LR is $\frac{\gamma}{\gamma^2}$.

So remarkably, while this is a 'specific-source' problem, and Mister X is identified, we computed the LR by using a common-source sampling model, since the difference between common and specific-source sampling models is exactly the extra hierarchical level added in a).

This observation led me to think some more about the application of common-source sampling models to specific-source problems. The application of common-source sampling models to specific-source problems might work for all feature-based LRs, since one can always add this extra hierarchical level and assume the distribution describing the characteristics of the population of potential reference sources known. However, does this approach (using common-source feature-based sampling models for specific-source problems) really make sense? And is there an analogous reasoning possible for score-based LRs, resulting in common-source score-based LRs being useful for specific-source problems? And if all these approaches turn out to be useful, which approach should we prefer?

### 1.4. Contributions of this paper

This paper makes contributions to the existing literature on five occasions.

1. In Section 2.4 the formal definition of all classes of LR systems studied in this paper results in a comprehensive overview of the relation between all these classes of LR systems in terms of a description by their sampling models. This also shows that specific-source and common-source LR systems that are otherwise of the same class are closely related when used for the trace-reference problem.

2. These definitions of LR systems imply that for some of the classes of LR systems, one has to be careful to properly condition the prior odds, as shown in Section 2.4.3.

3. When prior odds are properly conditioned, posterior odds can be properly factorized in likelihood ratio terms and prior odds, and one can study the performance of the several classes of LR systems by using strictly proper scoring rules (which operate on posterior probabilities). This allows for a ranking of the several classes of LR systems with respect to their benefit in answering specific-source (i.e. trace-reference comparison) questions, which is shown in Section 3.1.6.

4. In order to complement this theoretical analysis, the experimental burden required to realise these classes of LR systems is studied as well, using the description of their sampling models. This is done in Section 3.3.

5. Additionally, the present paper assesses whether the criticisms against the use of certain classes of LR systems hold ground from the perspective of evaluating performance by strictly proper scoring rules of the studied LR systems for the trace-reference problem, in Section 4.8.

The outline of the paper is as follows. Chapter 2 is concerned with a formal definition of the classes of LR systems studied. In the end of that chapter, I diverge a bit on the proper conditioning of anchored score-based LR systems, since the later chapters rely on properly conditioned LR systems and posterior probabilities. Chapter 3 ranks the different classes of LR systems in terms of performance, using so-called strictly proper scoring rules (Brümmer and Du Preez, 2006) as performance metrics and discusses practical feasibility when developing these systems. Chapter 4 examines the results in the context of existing literature and discusses implications of these results for future development of LR systems. In Appendix 2, strictly proper scoring rules are further introduced for readers who are not familiar with them.

## 2. Definitions and perspective of the paper

### 2.1. Assumptions

1. This paper addresses likelihood ratios. There is no integration over prior measures of belief, so Bayes factors are not considered.
2. In order to be able to compare classes of LR systems, the experimental design is kept the same. Thus, the objects (and sources they were taken from) in the object collections are constant for the different classes of LR systems. The partition of measurements over the objects is also constant, and the measurement results are also assumed the same.
3. The source from the crime scene is crime-related. If crime-relatedness of this source is disputed, one should not use source-level hypotheses but should revert to activity-level hypotheses (Cook et al., 1998).
4. All LR systems and probabilities for hypotheses discussed in this paper are conditioned on the same background information (relevant case circumstances and other measurements in the case that were already taken into account), but this conditioning term is omitted for readability.
5. All LR systems are well-calibrated, meaning that distributions reflect 'real-world' models.

### 2.2. Reframing specific and common-source scenarios for the trace-reference problem

Next, some reframing of terms is needed for more clarity. First, an 'inference of source where a specific source is known' is reframed from a 'specific-source' problem to a 'trace-reference' problem, since this paper studies (among other things) the merits of common-source sampling models to a problem where the reference can be identified. Trace-trace problems are out of scope. Furthermore, I need a redefinition of hypotheses used by Ommen and Saunders, since for example the notion for the common-source hypothesis pair that the source is unknown under $H_1$ does not apply to a trace-reference problem. Although this asks for a redefinition of the hypotheses, sampling models remain largely unchanged (in particular the use of the extra hierarchical level for common-source sampling models). When using specific-source sampling models, the only information about the reference objects comes from measurements on the reference (and the trace under $H_1$). When using common-source models, information about the reference objects also comes from measurements on a relevant background population of references.

This brings me to the following, more useful, definitions for studying the trace-reference problem:

Hypotheses, applicable to the trace-reference problem:

$H_1$: $o_r$ and $o_t$ both originate from the same source. This source can be identified.

$H_2$: $o_t$ originates from a different source than the source of $o_r$. The source of $o_r$ can be identified.

And sampling models, specific source:

$M_1$: $o_t$ is generated by random selection from the population of objects that can be generated from the specific source.

$M_2$: $o_t$ is generated by first randomly selecting a new source from the same population as used in $M_a$ (see below) and then randomly selecting objects from within that source.

$M_r$: $o_r$ is a collection of objects generated by random selection from the population of objects that can be generated from the specific source.

$M_a$: $o_a$ is a collection of objects generated by first randomly selecting $n_a$ sources from a relevant background population of sources and then randomly selecting $n_i$ objects from within the i-th source, for $i = 1, \ldots, n_a$.

And sampling models, common source:

$M_1$: $o_r$ and $o_t$ are generated by first randomly selecting a source from the population of crime-relevant sources (as also used in $M_{a1}$, see below) and then randomly selecting objects from within that source.

It may seem counterintuitive that $M_1$ concerns a randomly selected source, and not the specific source that was identified. However, *as far as we know*, this specific source may be regarded as selected randomly from this population.

$M_2$: $o_r$ is generated by first randomly selecting a source from a relevant background population of sources and then randomly selecting objects within that source. Next, $o_t$ is generated by first randomly selecting a source from a different relevant population of sources and then randomly selecting objects within that source.

$M_{a1}$: $o_{a1}$ is a collection of objects generated by first randomly selecting $n_{a1}$ sources from a population of crime-relevant sources and then randomly selecting $n_i$ objects from within the i-th source, for $i = 1, \ldots, n_{a1}$.

$M_{a2}$: $o_{a2}$ is a collection of objects generated by first randomly selecting $n_{a2}$ sources from a population of crime-irrelevant sources and then randomly selecting $n_j$ objects from within the j-th source, for $j = 1, \ldots, n_{a2}$.

Both collections of objects for $M_{a1}$ and $M_{a2}$ are used to represent a distribution of objects from sources. Under $H_1$ the source is crime-relevant by definition, and therefore the population of sources that are addressed by $M_1$ and $M_{a1}$ are the same. Under $H_2$ however, either the trace or reference objects are crime-irrelevant (depending on whether the trace or reference object is from the crime scene) and then one uses $M_{a2}$ to represent the crime-irrelevant population of sources. For example, think of a glass trace object. When the crime concerns a broken glass window from a burglary, a crime-relevant population of sources under $H_1$ may be other glass windows, while a relevant population of glass sources under $H_2$ may be table glass (the crime-irrelevant sources).

### 2.3. Other symbols used in this paper

In the paper, lower case letters are variables that have a fixed value, upper case letters are used for random variables. Hypotheses and sampling models have already been defined in Section 2.2.

Other symbols used:

$x \coloneqq$ features/measurement results on trace objects

$y \coloneqq$ features/measurement results on reference objects

$E \coloneqq \{X, Y\}$, features of trace and reference objects

$r \coloneqq$ specific reference source with its parameter values

$H_{As} \coloneqq$ hypothesis for the trace-reference problem, assuming specific-source sampling models,

and

$H_{Ac} \coloneqq$ hypothesis for the trace-reference problem, assuming common-source sampling models,

where

$A \coloneqq$ random variable taking values {1,2} in $H_{As}$ or $H_{Ac}$

$\delta(x, y) \coloneqq$ function of $x$ and $y$. Also called a 'score', a scalar value. $\Delta(X, Y) \coloneqq$ random variable equivalent.

$P(..) \coloneqq$ probability

$p(..) \coloneqq$ probability density

$f(..) \coloneqq p(..)$ or $P(..)$

$...^i \coloneqq$ index (superscript) assigned to randomly assigned value when sampling from a distribution, as in for example: $o_t^i$ (which can have many possible values, including $o_t$).

Furthermore, I use the following populations of sources to sample from:

C := background population of crime-relevant sources.

D := background population of reference sources given $H_2$. If the reference is from the crime scene, it is the same as C.

T := background population of trace sources given $H_2$. If the trace is from the crime scene, it is the same as C.

Using the assumption that the source or object found at the crime scene is crime-related, C is relevant for $H_{1c}$. D and/or T are relevant for $H_2$. Either T or D equals C, and the other one is a population of crime-irrelevant sources. If the distinction between crime-relevant and crime-irrelevant sources is irrelevant (think of, for example, the distribution of skin patterns on fingers or DNA profiles over perpetrators versus innocent people) then C = D = T.

## 2.4. Formal definition of LR systems

### 2.4.1. Feature-based LR systems

Below, I will define conditioning terms by referring to the hypothesis and sampling models specified in Sections 2.2 and 2.3.

For the definition of the specific-source feature-based LR, the paper of Neumann et al. (Neumann and Ausdemore, 2020) is followed.

**Specific-source**

$$SSFLR \coloneqq \frac{f(x,y|H_{1s})}{f(x,y|H_{2s})},$$

To make this definition less abstract, it may help to think about this in terms of 'the number of ways to sample evidence X and Y, by examining the generative models specified by the sampling and measurement models that come along with $H_{1s}$ and $H_{2s}$. One particular try *i* to obtain a particular $x^i$ and $y^i$ by sampling from these models is called a 'path'. The relative number of paths through which the observed evidence $x$ and $y$ can be obtained (as compared to all paths generating X and Y) is the desired probability (McElreath, 2020)[1].

Doing so, the probability in the numerator may be evaluated as: sample $o_t^i$ and $o_r^i$ from $r$. (I use the index '*i*' here to denote that $o_t^i$ and $o_r^i$ have one particular of many possible 'values', one of these possible values is $o_t$ and $o_r$.) After obtaining $o_t^i$ and $o_r^i$ by sampling from $r$, next sample $x^i$ and $y^i$ from $o_t^i$ and $o_r^i$ respectively. Do all the sampling steps over and over again, for $i = 1, ..., n$, where $n$ becomes so large the law of large numbers applies to a good approximation. Count the relative frequency that one sees $\{x, y\}$. The denominator can be evaluated as follows: sample $t^i$ from T, sample $o_t^i$ from $t^i$ and sample $o_r^i$ from $r$, sample $x^i$ and $y^i$ from $o_t^i$ and $o_r^i$ respectively. Do all the sampling steps over and over again, for $i = 1, ..., m$, where $m$ becomes so large the law of large numbers applies to a good approximation. Count the relative frequency that one sees $\{x, y\}$. The ratio of the two relative frequencies is the LR.

---

[1] A relative number of paths may actually only be defined for categorical data. However, for continuous data, a relative number of paths can be defined in a limiting form when defining bins for evidence ranges and letting the bin width approach zero. Then, the path analysis above also holds for continuous data.

According to the definitions below of the different classes of LR systems and their accompanying sampling paths, the listed LR systems have a lot of sampling steps in common. Figure 1 depicts the relations in terms of sampling paths for the different classes of LR systems studied in this paper. From this figure, it can for example be seen that specific-source and common-source LRs of the same class are related. For a specific-source LR, $r$ is treated as fixed, while for a common-cource LR it is treated as random. In the sampling steps that follow however, common-source and specific-source LR systems follow exactly the same paths[2]. Therefore, the only difference between common and specific-source LR systems of the same class is whether $r$ is treated as fixed or as random. This result will be used in Section 3.1.4 when studying the applicability of common-source sampling models to the trace-reference problem.

Fixing $r$ for the specific-source sampling models suits two different situations: $r$ is fixed and known (as for feature-based specific-source LR systems, see Neumann et al., 2020; Neumann and Ausdemore, 2020; Ommen and Saunders, 2018) and $r$ is fixed and unknown but we are able to model the distribution of the evidence as a function of $r$ (as for score-based specific-source LR systems). For common-source LR systems $r$ is treated as a random variable. This suits to a situation where we treat $r$ as unknown, and model the distribution of the evidence including uncertainty about $r$. This is the case for all common-source LR systems.

**Common-source**

$$CSFLR \coloneqq \frac{f(x, y | H_{1c})}{f(x, y | H_{2c})}$$

---

[2] For the SSXLR, this is not obvious from Figure 1. In the specific-source path, sampling $o_t^i$ from $r$ and subsequently sampling $x^i$ from $o_t^i$ under $H_1$ is omitted. One could however include these sampling steps, making common-source and specific-source LR paths identical up to sampling $r^i$ in the first step for a common-source LR system. The particular sampling steps were omitted in the Figure since it does not influence the specific-source LR; they are redundant. For SSYLR an analogous reasoning holds.

Let us do the 'path analysis' again. Numerator: sample $r^i$ from C, sample $o_t^{\ i}$ and $o_r^{\ i}$ from $r^i$, sample $x^i$ and $y^i$ from $o_t^{\ i}$ and $o_r^{\ i}$ respectively. Do all the sampling steps over and over again. Count the relative frequency that one sees $\{x, y\}$. Denominator: sample $t^i$ from T, and $r^i$ from D, sample $o_t^{\ i}$ from $t^i$ and $o_r^{\ i}$ from $r^i$, $x^i$ and $y^i$ from $o_t^{\ i}$ and $o_r^{\ i}$ respectively. Do all the sampling steps over and over again. Count the relative frequency that one sees $\{x, y\}$. Compared to an SSFLR, it has one more sampling step: sample $r^i$ from C (numerator) or D (denominator).

### 2.4.2. Score-based LR systems

In many studies score-based LR systems are used (Bolck et al., 2009; Hepler et al., 2012; Leegwater et al., 2017; Vergeer et al., 2014). These are most often used in situations where the probability distribution of the features is hard to come by. The reason for this is, for example, because the feature-space is very high dimensional and so the experimental demands to get an accordingly large enough sample size are too high, or when the probability models for features are hard to describe parametrically. In both of these cases a proposed solution is to calculate a scalar that can be interpreted as a comparison or difference score between $x$ and $y$. Such a function is defined as: $\delta(x, y)$.

#### 2.4.2.1. Non-anchored

LR systems are non-anchored when neither the measurement results of the reference nor of the trace objects are considered fixed. Non-anchored score-based LRs are defined as:

**Specific-source non-anchored**

$$SSSLR \coloneqq \frac{f(\delta(x,y)|H_{1s})}{f(\delta(x,y)|H_{2s})},$$

See Figure 1 and the path analysis is in Appendix 1.

**Common-source non-anchored**

$$CSSLR := \frac{f(\delta(x,y)|H_{1c})}{f(\delta(x,y)|H_{2c})}.$$

See Figure 1 and the path analysis is in Appendix 1.

### *2.4.2.2. Anchored*

In some studies anchoring for score-based LRs is proposed (Alberink et al., 2014; Hepler et al., 2012). For such LR systems, the measurement *y* on the reference objects or the measurement *x* on the trace objects in the case at hand is regarded as fixed. Y-anchored LR systems are defined as:

**Specific-source y-anchored**

$$SSYASLR := \frac{f(\delta(x,y)|H_{1s}, y)}{f(\delta(x,y)|H_{2s}, y)},$$

See Figure 1 and the path analysis is in Appendix 1.

**Common-source y-anchored**

$$CSYASLR := \frac{f(\delta(x,y)|H_{1c}, y)}{f(\delta(x,y)|H_{2c}, y)}.$$

See Figure 1 and the path analysis is in Appendix 1.

Finally, one may regard X-anchored LR systems that are conditioned on the measurement of the trace objects in the case. For the specific-source sampling models, it is defined as:

**Specific-source x-anchored**

$$SSXASLR := \frac{f(\delta(x,y)|H_{1s}, x)}{f(\delta(x,y)|H_{2s}, x)},$$

The LR of SSXASLR is 1 always (Neumann and Ausdemore, 2020). This can easily be inferred from the path analysis:

Numerator: sample $o_r^i$ from $r$, sample measurements $y^i$ from $o_r^i$. Calculate $\delta^i(x, y^i)$. Do all the sampling steps over and over again. Count the relative frequency that one sees $\delta$ in the paths. Denominator: sample $o_r^i$ from $r$, sample measurements $y^i$ from $o_r^i$. Calculate $\delta^i(x, y^i)$. Do all the sampling steps over and over again. Count the relative frequency that one sees $\delta$ in the paths. The two methods for path generation are identical, so the LR = 1 always. This is due to the fact that all paths concerning traces have to result in $x$ and only these paths could have been different for the numerator and the denominator. Since the LR is 1 always, this class of LR systems will not be considered further in this paper.

For the common-source sampling models, the situation is different since information about the reference characteristics also comes from measurements on some background population, and under $H_{1c}$ information about these characteristics also comes from $x$, whereas under $H_{2c}$ it does not, making the paths different under $H_{1c}$ then under $H_{2c}$:

**Common-source x-anchored**

$$CSXASLR := \frac{f(\delta(x,y)|H_{1c},x)}{f(\delta(x,y)|H_{2c},x)}.$$

Consider the path analysis. Numerator: sample $r^i$ from C, sample $o_t^i$ and $o_r^i$ from $r^i$, sample measurements $x^i$ and $y^i$ from $o_t^i$ and $o_r^i$ respectively. Keep all paths for which $x^i$ equals $x$. Calculate $\delta^i(x, y^i)$. Do all the sampling steps over and over again. Count the relative frequency that one sees $\delta$ in the kept paths. Denominator: sample $r^i$ from D, sample $o_r^i$ from $r^i$, sample $y^i$ from $o_r^i$. Calculate $\delta^i(x, y^i)$. Do all the sampling steps over and over again. Count the relative frequency that one sees $\delta$ in the paths. The two methods for path generation are different, so the LR is generally different from 1.

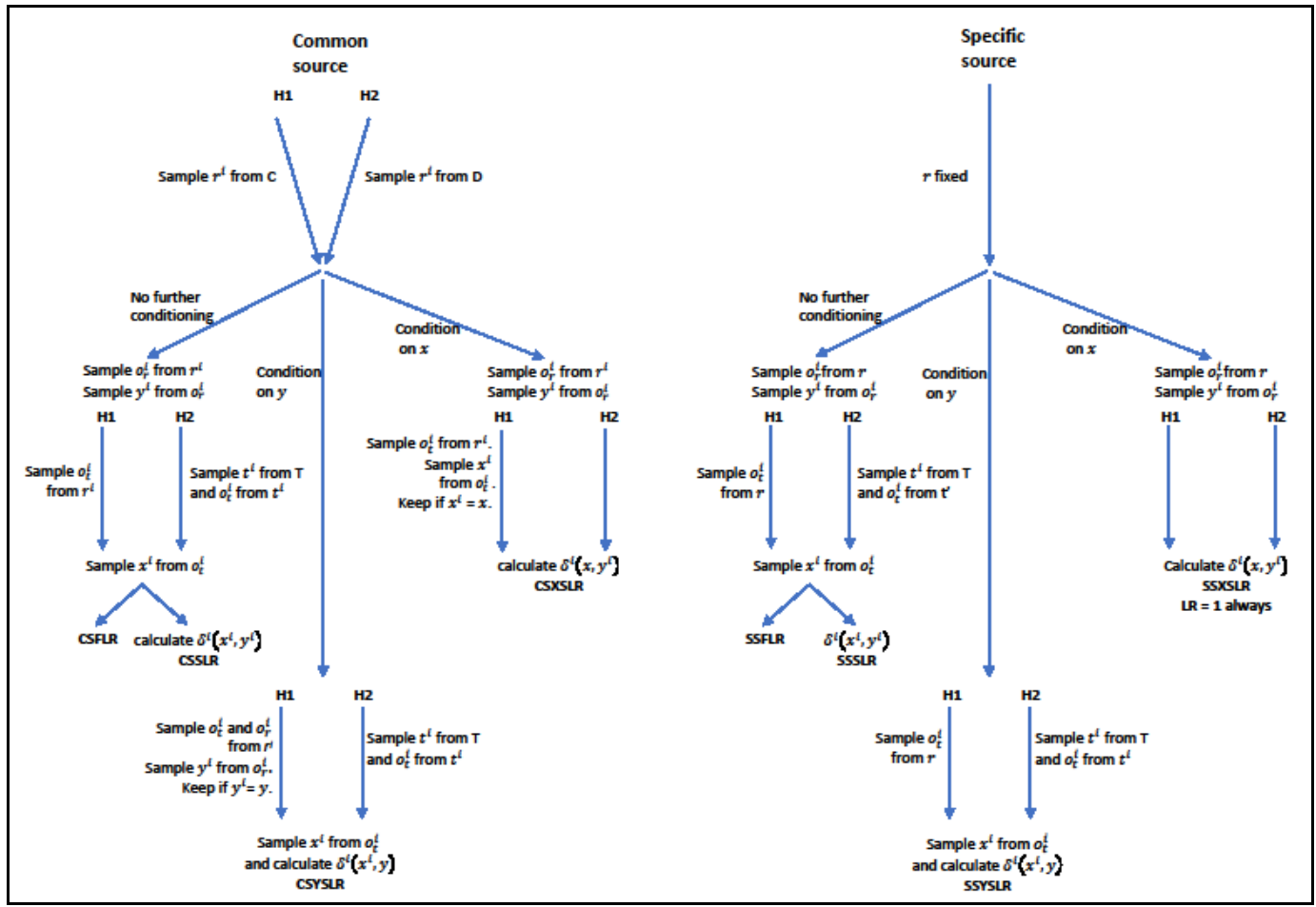

Figure 1. Path analysis: overview of classes of LR systems in terms of the sampling steps to generate their respective evidence.

### 2.4.3. Some anchored LR systems may be ill-conditioned

When anchoring certain classes of score-based likelihood ratio systems, one may mess up the rules for proper conditioning. This is the case for three of the LR systems mentioned above. To understand what is meant by messing up the conditioning, note the following.

Irrespective of common- or specific-source sampling models, the posterior odds of a $a$-anchored score $\delta$ (where $a$ may be $x$ or $y$) are given by:

$$\frac{f(H_1|\delta, a)}{f(H_2|\delta, a)}$$

By applying Bayes rule this results in:

$$\frac{P(H_1|\delta,a)}{P(H_2|\delta,a)} = \frac{f(\delta|H_1,a)}{f(\delta|H_2,a)} \times \frac{f(a|H_1)}{f(a|H_2)} \times \frac{P(H_1)}{P(H_2)}$$

The right-hand side of the equation has two likelihood ratio terms (and a prior odds term), and the right LR term may be different from one, depending on whether the paths to get to $a$ are the same or different under $H_1$ and $H_2$. The anchored score-based LR is given by the left LR term. When the right LR term is different from one, evaluating only the left LR term to update the prior odds for the hypotheses imposes improper conditioning.

Consider the $CSXASLR$. Paths are different for $a = x$ when C is different from T, since paths to $x^i$ go through $t^i$ and $t^i$ is sampled from C under $H_1$ and from T under $H_2$. Consider such a system for glass evidence, when the case is about a burglary in a house where a glass window is broken. In that case, the relevant background population under $H_1$ for glass trace fragments found on a suspects clothing may be window glass, while the relevant background population under $H_2$ may constitute table glass. Therefore, C is different from T. This means that $CSXASLR$ may be ill-conditioned.

Paths are also different for $a = y$ for $CSYASLR$ when C is different from D, since paths to $y^i$ go through $r^i$ and $r^i$ is sampled from C under $H_{1c}$ and from D under $H_{2c}$. For example, think of an investigation about whether a bullet was fired with a specific gun, in a hitman shooting incident and the gun is found in a trash container. The relevant population C of guns under $H_{1c}$ may be other guns typically used in such shooting incidents, while the relevant population D of guns under $H_{2c}$ may be a more general one.

When examining types of forensic evidence in general, a distinction may be made between whether the reference or the trace is from the crime scene. When the reference is from the crime scene (like in the glass example), the term $\frac{f(y|H_1)}{f(y|H_2)}$ equals 1, while the term $\frac{f(x|H_1)}{f(x|H_2)}$ may be different from 1.

When the trace is from the crime scene (like in the gun example) the term $\frac{f(x|H_1)}{f(x|H_2)}$ equals 1 while the term $\frac{f(y|H_1)}{f(y|H_2)}$ can be different from 1[3]. There are a few evidence modalities where both $\frac{f(x|H_1)}{f(x|H_2)}$ and $\frac{f(y|H_1)}{f(y|H_2)}$ are equal to 1, since for these evidence modalities C = D = T. An overview of some evidence modalities is given in Table 1.

| Trace crime-relevant[4] | Reference crime-relevant | Distinction not important[2] |
|---|---|---|
| $\frac{f(x\|H_1)}{f(x\|H_2)} = 1,$  $\frac{f(y\|H_1)}{f(y\|H_2)} = ?$  $C = T \neq D$ | $\frac{f(x\|H_{1c})}{f(x\|H_{2c})} = ?,$  $\frac{f(y\|H_{1c})}{f(y\|H_{2c})} = 1$  $C = D \neq T$ | $\frac{f(x\|H_1)}{f(x\|H_2)} = 1,$  $\frac{f(y\|H_1)}{f(y\|H_2)} = 1$  $C = D = T$ |
| Cartridge cases  Bullets  Fibers on victim | Glass  Soil  Fibers on suspect  Speech recognition | DNA  Finger prints  Voice recognition[5]  Face recognition[5] |

Table 1 Overview of some forensic evidence types in three categories distinguished by whether $\frac{f(x|H_1)}{f(x|H_2)}$ and/or $\frac{f(y|H_1)}{f(y|H_2)}$ are equal to 1.

From now on, I assume that LR systems are used that have proper conditioning. This means that when common-source anchored LR systems are used, the feature that is anchored on has an LR of 1. For the left column in Table 1, anchoring on *x* is appropriate, for the middle column anchoring on *y* is appropriate and for the right column anchoring on $x$ or $y$ is appropriate. Derivations in Section 3.1.2 do not hold for ill-conditioned anchored LR systems.

---

[3] Another way to put this: background information I is sometimes defined as information that in itself has no evidential value but is needed to properly define populations under $H_1$ and $H_2$. In this respect, these anchored SLR systems are properly conditioned when *x* or *y* can be seen as part of the background information.
[4] Sampling models can be either specific or common source.

[5] In voice and face recognition, the trace is crime-relevant, so $\frac{f(x|H_1)}{f(x|H_2)} = 1$. $\frac{f(y|H_1)}{f(y|H_2)}$ equals 1 also since usually for $f(y|H_{1c})$ and $f(y|H_{2c})$ the distribution is conditioned on some characteristics of the reference voice (for example dialect and male/female) or face (race), it is these characteristics that would distinguish the distributions $f(y|H_{1c})$ and $f(y|H_{2c})$. Therefore, these characteristics have evidential values themselves and it is assumed that the judiciary takes this evidential value into account in the a priori probabilities.

# 3. Ranking the stars

In this chapter, I rank the introduced classes of LR systems in terms of performance and in terms of experimental effort to realise them.

## 3.1. Ranking the stars by performance on strictly proper scoring rules

I will use strictly proper scoring rules to compare relative performance of the classes of LR systems. Scoring rules quantify the goodness of decisions, assuming that decisions are based on comparing a probability *P* for an event to some threshold value *Th* for deciding whether the event has occurred or not. When $P > Th$, it is decided that the event has occurred. When $P < Th$, it is decided that the event has not occurred. In court cases, the exact value of such a threshold in deciding between the occurrence of hypotheses is usually unknown, but this is mitigated since strictly proper scoring rules evaluate goodness of decisions over a distribution of thresholds (Brümmer and Du Preez, 2006). Each particular strictly proper scoring rule is defined by a different distribution over thresholds. This makes the class of strictly proper scoring rules very versatile when comparing relative performance of classes of LR systems. For readers unfamiliar with strictly proper scoring rules, see Appendix 2 for a more elaborate introduction.

When studying performance of LRs, two hypothesis $H_1$ and $H_2$ are contrasted. In that case, a strictly proper scoring rule has the following property (see Eq 5.2 in Degroot and Fienberg, 1983, and their text right below).

$$\sum_{a=\{1,2\}} P(H_a|e) SPSR[P(H_a|e)] \geq \sum_{a=\{1,2\}} P(H_a|e) SPSR[P(H_a|b)], \qquad \text{(Eq. 1)}$$

Where *e* and *b* are different conditioning information, and (for example) $SPSR[P(H_1|e)]$ and $SPSR[P(H_1|b)]$ are the values of the strictly proper scoring rule evaluated when $A = 1$. Equality in Eq. 1. is achieved only when $b = e$. Eq. 1. Means that using 'true' probabilities $P(H_a|e)$ for the SPSR to operate on results in maximal expected score. And the higher the expected score, the better the expected performance.

I will compare performance of LR systems when they are used repeatedly in court. I will do this for an imaginary list of court cases by taking averages of the SPSR-metrics over cases. Since varying cases means varying evidence, trace and references and the ground truth, averages should be taken over all these variables. Again, see Appendix 2 for a more elaborate justification. I take averages by using integrals representing expected values, and use a summation for the binary ground truth states. The evidence collection {X,Y} is represented by $E$, reference characteristics are represented by R.

The core of all derivations below is based on the law of total expectation. For any two random variables *K* and *M*, probability density functions *p(…)* and any function $F(m)$ it reads:

$$\int F(m)p(m)dm = \int \int F(m)p(k)p(m|k)dkdm$$

I use this formula, but then for the case when $M$ consists of two variables.

For example, in the conversion of Eq. 2 the law of total expectation is used, with

$$M = \{\Delta, H_A\}$$

$$K = E$$

$$F(M) = SPSR[P(H_A|\Delta)]$$

Therefore, I use a version of the law of total expectation where the number of variables described by $M$ is 2. For a more elaborate justification, see Appendix 3.

### 3.1.1. Feature-based LR systems versus score-based LR systems

Imagine comparing relative performance of these two systems over a sequence of court cases, by evaluating posterior probabilities, using these LR systems to update prior probabilities and using an SPSR as a performance metric. Conditioning information varies over these cases (with regard to the posterior probabilities, the evidenceand the specific reference source is different). Note that the a priori probability distribution $P(H_A)$ is constant since in section 2.1 (Assumption 4) it was assumed that the background information *I* was constant[6].

We prove that:

$$\int f(e) \sum_a P(H_a|e) SPSR[P(H_a|e)]\, de \geq \int f(\delta) \sum_a P(H_a|\delta) SPSR[P(H_a|\delta)]\, d\delta \qquad \text{(Eq. 2)}$$

Eq. 2 states that a score-based LR system (based on the same features) will never outperform a feature-based LR system, in terms of expected performance on strictly proper scoring rules.

Eq. 2 may be interpreted as follows. The left-hand side evaluates the expected value of an $SPSR$ for the posterior distribution of *H* given the features *E*, with respect to a sequence of court cases. The sequence of court cases is represented by a probability distribution over $\{H_A, E\}$ (over cases, *A* and *E* vary, and their variation is described by $f(e) \times P(H_a|e)$). The average performance is evaluated by taking the expected value of the SPRS with respect to $\{H_A, E\}$. $P(H_a|e)$ is the posterior probability combining the feature-based LR system with the prior probability. The right-hand side evaluates the expected value of an SPSR for a posterior distribution of *H* given score Δ, for the same sequence of court cases, but now with knowing the score Δ instead of *E*. $P(H_A|\Delta)$ is the posterior probability using the score-based LR system in combination with the prior probability.

We convert the right-hand side, by introducing *E* and using the law of total expectation:

$$\int f(\delta) \sum_a P(H_a|\delta) SPSR[P(H_a|\delta)]\, d\delta = \int f(e) \int f(\delta|e) \sum_a P(H_a|\delta, e) SPSR[P(H_a|\delta)]\, d\delta\, de$$

Since δ is a deterministic function of *e*, it holds that:

$$f(\delta|e) = g(\Delta = \delta(e)),$$

---

[6] All the following derivations hold also for varying conditioning information *I*. The effect of varying *I* is that all expectations (all derivations below concern the comparison of expected values) should also be taken with respect to *I*, averaging over different terms given *I*. Since all terms given *I* have the same inequality properties, the average also has this inequality property.

where $g$ is the Dirac delta function. Further note that H is conditionally independent of $\Delta$ given $E$ and thus $P(H_a|\delta, e) = P(H_a|e)$.

Therefore, the right-hand side of Eq. 2 can be written as:

$$= \int f(e) \sum_a P(H_a|e) SPSR[P(H_a|\delta(e))] de$$

Note that $P(H_a|\delta(e))$ can be evaluated when $E$ is known (since $\Delta$ is a deterministic function of $E$).

Eq 2 becomes:

$$\int f(e) \sum_a P(H_a|e) SPSR[P(H_a|e)] \, de \geq \int f(e) \sum_a P(H_a|e) SPSR[P(H_a|\delta(e))] de,$$

Note that for all $e \in E$:

$$\sum_a P(H_a|e) SPSR[P(H_A|e)] \geq \sum_a P(H_a|e) SPSR[P(H_a|\delta(e))]$$

by definition of using a strictly proper scoring rule, which proves the claim.

### 3.1.2. Anchored score-based LR systems versus non-anchored score-based LR systems

The relation between an anchored and a non-anchored LR system is the fact that the anchored LR system is further conditioned as compared to the non-anchored LR system. Below, it will be shown that, keeping hypotheses the same, any further conditioning with some evidence $Z = X$ or $Y$ cannot lead to worse performance.

We prove the following:

$$\int f(\delta) \int f(z|\delta) \sum_a P(H_a|z, \delta) SPSR[P(H_a|z, \delta)] dz d\delta \geq \int f(\delta) \sum_a P(H_a|\delta) SPSR[P(H_a|\delta)] \, d\delta$$

(Eq. 3)

The left-hand side of Eq. 3 evaluates a strictly proper scoring rule where conditioning on $Z$ and $\Delta$ is applied. It is the expected value over $Z$ and $\Delta$ of an SPSR for posterior probability distribution given $Z$ and $\Delta$ (so for the anchored score-based LR system). The right-hand side is the expected value of an SPSR for the probability distribution for the hypotheses not conditioned on $Z$ (so for the non-anchored score-based LR system).

On the right-hand side of the inequality, $Z$ is introduced by the law of total expectation. This transforms Eq. 3 to:

$$\int f(\delta) \int f(z|\delta) \sum_a P(H_a|z,\delta) SPSR[P(H_a|z,\delta)] \, dz d\delta$$

$$\geq \int f(\delta) \int f(z|\delta) \sum_a P(H_a|z,\delta) SPSR[P(H_a|\delta)] \, dz d\delta$$

Which proves the claim, since for all for all $\{z,\delta\} \in \{Z,\Delta\}$:

$$\sum_a P(H_a|z,\delta) SPSR[P(H_a|z,\delta)] \geq \sum_a P(H_a|z,\delta) SPSR[P(H_a|\delta)],$$

By definition of using an SPSR.

Note that this derivation assumes that anchored LR systems do not suffer from being ill-conditioned as explained in Section 2.4.3. If anchored LR systems are ill-conditioned, the derivation does not hold, since $P(H_a|z,\delta)$ is not well-defined.

### 3.1.3. Comparing the use an LR system versus using prior odds only

It will be shown that using any LR system (score-based, feature-based, anchored, non-anchored), cannot perform worse than using the prior odds only.

We prove the following:

$$\int f(v) \sum_a P(H_a|v) SPSR[P(H_a|v)] dv \geq \sum_a P(H_a) SPSR[P(H_a)],$$

Where $V$ is any further conditioning. For example, $V$ is $\{X, Y\}$ or $\Delta$.

Note that the left-hand side gives the expected performance when using an LR system to update the prior odds and the right-hand side gives the expected performance when using the prior probabilities only.

Note that by introducing $V$ to the right-hand side by the law of total expectation the inequality becomes:

$$\int f(v) \sum_a P(H_a|v) SPSR[P(H_a|v)] dv \geq \int f(v) \sum_a P(H_a|v) SPSR[P(H_a)] dv$$

and the proof is immediate from the definition of an SPSR.

### 3.1.4. Common-source LR systems for the trace-reference problem

In forensic practice, many LR systems that are used in casework for the trace-reference problem are, in fact, common-source LR systems. Therefore, it is worthwhile to study this situation. Below, it will be shown that common-source sampling models are applicable to the trace-reference problem.

As a start, note that all common-source likelihoods can be seen as averages of specific-source likelihoods, using a prior distribution for $r$. We therefore introduce $r$ as a nuisance parameter in the common-source likelihood:

$$f(v|H_{ac}) = \int f(v|r, H_{ac}) f(r|H_{ac}) dr = \int f(v|H_{as}) f(r|H_{ac}) dr \qquad \text{(Eq. 4)}$$

Where $v$ is either a score or the features. The equality of $f(v|r, H_{ac})$ and $f(v|H_{as})$ can be seen from the path analysis depicted in Figure 1. Consider a common-source LR where $r$ is known. This transforms the common-source LR to a specific-source LR: all sampling paths for $r^i$ result in $r$ and then the common and specific-source LRs are identical. The right hand side of Eq. 4 expresses the common-source likelihood as an average of specific-source likelihoods, with as prior distribution for $r : f(r|H_{ac})$.

Eq. 4 treats the uncertainty in $R$ by integrating over it. This is perfectly in line with Bayesian probability calculus, where it is prescribed to integrate over uncertainty in nuisance parameters this way. $r$ can be seen as a nuisance parameter, since it is fixed and unknown. Therefore, from a Bayesian perspective common-source LRs are perfectly valid for the trace-reference problem.

From this perspective, in a $CSFLR$ the prior distribution on $r$ is updated by $y$, the measurements on the specific source. In analogy, for a $CSYASLR$ the prior distribution on $r$ is updated by $y$ as well, and for a $CSSLR$ only the prior distribution for $r$ is used as information for $r$.

The specific reference is seen as a random sample from this prior distribution for $r$. The prior distribution is then specified by a suitable choice of a population of references under $H_{Ac}$. According to the path analysis from Section 2.4 a suitable choice for this population is C when $H_{Ac}= H_{1c}$ and D when $H_{Ac}= H_{2c}$.

Finally, note that a limiting case for a common-source LR is when the parameters of the reference are known a priori, resulting in a specific-source LR.

The reasoning above shows that, from the viewpoint of Bayesian probability calculus, common-source LRs integrate out parameter uncertainty concerning $r$, and common-source sampling models are valid approaches to the trace-reference problem. When doing so, it is assumed no information about $r$ is incorporated in the prior probabilities for the hypotheses.

However, when studying the literature on specific-source likelihood ratios (Neumann et al., 2020; Neumann and Ausdemore, 2020; Ommen and Saunders, 2018), these works may suggest that for specific-source priors for hypotheses, one could also condition those prior probabilities on $r$. In these studies it is assumed that $r$ is not only fixed, but also known, and $r$ therefore may be used as background information for the specific-source LRs. This means that when using a specific-source sampling model the prior probability for the hypotheses is already conditional on $r$. This makes specific-source prior odds different from common-source prior odds, where for the latter no conditioning on $r$ is used. Therefore, in the sections that follow, we study the situation where specific-source prior odds are conditioned on $r$, and these are updated by a common-source LR.

In that situation, for most evidence modalities it turns out that updating a specific-source prior with a common-source LR also leads to an expected performance that cannot be worse than using only the specific-source prior, under one condition. The condition is that $\frac{f(r|H_{1c})}{f(r|H_{2c})} = 1$. In Sections 3.1.4.1 to 3.1.4.3 this result is derived.

This may be an intriguing result, since updating such specific-source prior odds with a common-source LR is a very strange situation at first sight. It seems a violation of Bayes' rule, since the background information is different for the prior odds and the LR (the prior odds are conditioned on $r$ while the common-source likelihood ratio is not conditioned on $r$). However, as long as $\frac{f(r|H_{1c})}{f(r|H_{2c})}$ turns out to be 1, the conditioning of the prior odds on $r$ is irrelevant.

### 3.1.4.1. Relation between common-source and specific-source posterior probabilities

Since common- and specific-source likelihood functions are the same when $r$ is given (see text below Eq. 4), it follows that their posterior probabilities are also the same when $r$ is given

$$P(H_{as}|v,r) = P(H_{ac}|v,r), \qquad (Eq.\ 5)$$

for any $V$, including the empty set. Eq. 5 is true since hypotheses are the same, and sampling models for the LR of $v$ given $r$ are also the same.

We are going to use Eq. 5 for the proof below.

### 3.1.4.2. The claim

We have to prove:

$\int f(r) \int f(v|r) \sum_a P(H_{as}|v,r) SPSR[P(H_a|v_{cs},r)] \, dv \, dr \geq$

$\int f(r) \sum_a P(H_a|r) SPSR[P(H_a|r)] \, dr$, (Eq. 6).

The right-hand side of Eq. 6 gives expected performance for the prior conditioned on $R$. The left-hand side of Eq. 6. gives expected performance of a prior conditioned on $R$ updated by a common-source LR, denoted by: $P(H_a|v_{cs},r)$, in light of a specific-source posterior probability. The odds equivalent of $P(H_a|v_{cs},r)$ is:

$$\frac{P(H_1|v_{cs},r)}{P(H_2|v_{cs},r)} = \frac{f(v|H_{1c})}{f(v|H_{2c})} \times \frac{P(H_1|r)}{P(H_2|r)},$$ (Eq. 7)

Clearly, this is a violation of Bayes' rule. But this is exactly what happens when one updates prior odds that are conditioned on $R$ with a common-source LR this way. The question is: how detrimental is the loss in performance when we do so? Let us compare it versus not updating this prior, as described by the right-hand side of Eq. 6.

In the reformulation of Eq. 6 that follows below, the goal is to integrate $R$ out of the equation, and to write the equation in terms of common-source hypotheses/sampling models, using the relations between the common and specific-source likelihoods and posterior probabilities defined in Eq. 4 and Eq. 5.

Substituting $P(H_{As}|r) = P(H_{Ac}|r)$ and $P(H_{As}|v,r) = P(H_{Ac}|v,r)$ (see Eq. 5) in Eq. 6 we get an equivalent claim:

$\int f(r) \int f(v|r) \sum_a P(H_{ac}|v,r) SPSR[P(H_a|v_{cs},r)] \, dv \, dr \geq$

$\int f(r) \sum_a P(H_a|r) SPSR[P(H_a|r)] \, dr$, (Eq. 8).

### 3.1.4.3. *Proving the claim*

Using the claim in Eq. 8, the proof goes as follows. First, we assume that $P(H_a|r)$ is independent of $R$:

$$P(H_a|r) = P(H_a),$$

For $r \epsilon R$.

The implications of this assumption will be discussed after proving the claim.

Now note that since $P(H_a|r) = P(H_a)$ it also follows that $P(H_a|v_{cs}, r)$ is independent of $R$: Eq. 7 right-hand side is:

$$\frac{P(H_1|v_{cs}, r)}{P(H_2|v_{cs}, r)} = \frac{f(v|H_{1c})}{f(v|H_{2c})} \times \frac{P(H_1|r)}{P(H_2|r)},$$

and following the assumption $P(H_a|r) = P(H_a)$, there is no dependence on $r$ anymore. Then one can write $P(H_{ac}|v)$ for $P(H_a|v_{cs}, r)$. Doing this substitution, the left-hand side of Eq. 8 becomes:

$$\int f(r) \int f(v|r) \sum_a P(H_{ac}|v, r) SPSR[P(H_{ac}|v)] \, dv dr$$

By shuffling $R$, $V$ and $H_{Ac}$ using laws of probability this is equal to:

$$\sum_a P(H_{ac}) \int SPSR[P(H_{ac}|v)] f(v|H_{ac}) \int f(r|v, H_{ac}) dr dv, \qquad \text{(eq. 9)}$$

By integrating $R$ out, Eq. 9 can be written as:

$$= \sum_a P(H_{ac}) \int SPSR[P(H_{ac}|v)] f(v|H_{ac}) \, dv.$$

for $r \in R|H_{Ac}$.

Writing $f(v|H_{ac}) = \frac{P(H_{ac}|v) f(v)}{P(H_{ac})}$, this results in:

$$= \int f(v) \sum_a P(H_{ac}|v) SPSR[P(H_{ac}|v)] \, dv$$

Now consider the right-hand side of Eq. 8. Introducing $V$ and integrating $R$ out analogous to the derivation above gives

$$Eq\ 8, RHS = \int f(r) \int f(v|r) \sum_a P(H_{as}|v,r) SPSR[P(H_a|r)]\, dv dr$$

$$= \int f(v) \sum_a P(H_{ac}|v) SPSR[P(H_a)]\, dv$$

for $r \in R|H_{Ac}$.

Substituting both left and right terms in Eq. 8:

$$\int f(v) \sum_a P(H_{ac}|v) SPSR[P(H_{ac}|v)]\, dv \geq \int f(v) \sum_a P(H_{ac}|v) SPSR[P(H_a)]\, dv,$$

for $r \in R|H_{Ac}$

Which proves the claim since the inequality holds for all $\sum_a (\dots)$ terms.

Finally, let us discuss the implications of the assumption $(H_a|R) = P(H_a)$.

The corresponding odds can be factorised in:

$$\frac{P(H_1|r)}{P(H_2|r)} = \frac{f(r|H_1)}{f(r|H_2)} \times \frac{P(H_1)}{P(H_2)}, \qquad \text{(Eq. 10)}$$

for any $r \in R|H_A$.

Since the prior odds $\frac{P(H_1)}{P(H_2)}$ do not depend on $r$, the term $P(H_A|r)$ is independent of $r$ when the likelihood ratio $\frac{f(r|H_1)}{f(r|H_2)}$ equals 1 for any $r$. In Section 2.4.3 it was already discussed for which evidence types this is the case. It holds for evidence types where the specific source is from the crime scene or when the distinction 'trace is from crime scene' versus 'reference is from crime scene' is not important (columns 2 and 3 in Table 1).

When the trace is from the crime scene however (column 1 in Table 1), the assumption does not hold. In this situation, the prior probability for the hypotheses does depend on $r$ when conditioning on it. In that case, the trier of fact should evaluate $\frac{f(r|H_1)}{f(r|H_2)}$ when evaluating the prior odds. This seems strange, since the forensic expert is being asked to help interpreting measurements made on $r$ (by either providing a common- or a specific-source LR), which are then also needed to get information relevant to $\frac{f(r|H_1)}{f(r|H_2)}$.

This seems to be a complication when assuming $r$ is known as background information: the evidential value of $r$ is hidden in the prior odds. A common-source approach would not suffer from this shortcoming, since $r$ is not used as background information; one never conditions on $r$. A specific-source LR when assuming $r$ is known as background information would suffer from this shortcoming. Assuming the trier of fact is aware nor able to evaluate $\frac{f(r|H_1)}{f(r|H_2)}$ it is unclear for this particular situation (when $\frac{f(r|H_1)}{f(r|H_2)}$ is not equal to one) which sampling model - specific or common source - is the optimal approach to this problem. They both have their particular drawback: common-source LRs generally perform less than specific-source LRs (see below), but assuming $r$ is known as background information for a specific-source LR when there is evidential value in $r$ itself also results in decreased performance .

### 3.1.5. *Specific-source LR systems versus common-source LR systems*

Assuming $\frac{f(r|H_1)}{f(r|H_2)}$ is evaluated correctly and keeping the evidence the same, it is shown that a specific-source LR system cannot perform worse than a common-source LR system, when used to update prior odds assuming $r$ is known a priori (and properly conditioned), and assuming $r$ is not known a priori.

When $r$ is known a priori we prove that:

$$\int f(r) \int f(v|r) \sum_a P(H_{as}|v,r) SPSR[P(H_{as}|v,r)] \, dv dr$$

$$\geq \int f(r) \int f(v|r) \sum_a P(H_{as}|v,r) SPSR[P(H_{as}|v_{cs},r)] \, dv dr,$$

where $V$ is any further conditioning.

The proof is immediate since $P(H_{as}|v,r) \neq P(H_{as}|v_{cs},r)$.

When $r$ is not known a priori we prove that:

$$\int f(v) \sum_a P(H_{as}|v) SPSR[P(H_{as}|v)] \, dv \geq \int f(v) \sum_a P(H_{as}|v) SPSR[P(H_{ac}|v)] \, dv$$

The proof is immediate since $P(H_{as}|v) \neq P(H_{ac}|v)$.

### 3.1.6. Overview of performance results

An overview of the performance results derived in the sections above is given below, with reference to the section where the result is obtained. A legend for the abbreviations used is given in the Table caption.

| S3.1.1 | S3.1.2 | S3.1.3 | |
|---|---|---|---|
| $SSFLR \geq$ | $SSYASLR \geq$ | $SSSLR \geq$ | prior |

| S3.1.1 | S3.1.2 | S3.1.3 | |
|---|---|---|---|
| $CSFLR \geq$ | $\begin{array}{c}CSYASLR\\CSXASLR\end{array} \geq$ | $CSSLR \geq$ | prior |

| S3.1.4 |
|---|
| $CSLR \geq prior^*$ |

| S3.1.5 |
|---|
| $SSSLR \geq CSSLR$ |
| $SSYASLR \geq CSYASLR$ |
| $SSFLR \geq CSFLR$ |

Table 2. Overview of performance results (comparing posterior probabilities for hypotheses updated using a certain class of LR system). Abbreviations used: SS = specific-source, CS = common-source, S = score-based, F = feature-based, A

= anchored, Y = features of reference, X = features of trace. In row 3, the term $CSLR$ denotes any common-source LR system. * When $r$ is assumed known a priori this holds when $\frac{f(r|H_1)}{f(r|H_2)} = 1$.

## 3.2. Practical feasibility

Practical feasibility will be assessed by how much experimental effort is needed when developing an LR system belonging to a certain class. Thus, I assume that experiments are needed to get data[7]. For the different classes of LR systems it will be identified what typical requirements are for the number of measurements on the trace and reference objects in the case and for measurements on other objects from sources from background populations. Where the former section was theoretical and exact in nature, the following section is partly based on personal experience of the author. In addition, the required experimental effort in casework depends on the a priori assessment of the possible magnitude of the likelihood ratio for the comparison at hand. In general, if a more discriminating LR value is expected, one also needs more measurements to substantiate such a value. Therefore, in the following section general trends in required number of measurements are more important than the exact numbers. In the section that follows, I assume that the goal is to calculate LRs from roughly 1/100 to 1000. As a rule of thumb[8] to achieve this for score-based LR systems, 1000 scores under $H_2$ are needed and 100 scores under $H_1$. The asymmetry in smallest and largest LR value comes about since, as it turns out, it is generally experimentally more easy to get scores under $H_2$ than under $H_1$ for a score-based LR system.

### 3.2.1. Specific-source versus common-source LR systems

When assessing practical feasibility, it is helpful to look at the path analyses in Section 2.4 and Appendix 1 again. From the path analysis, it becomes clear what sample of sources is needed to sample objects from (and the objects are subsequently measured). For example, an important distinction between paths for specific-source and common-source LR systems is that for specific-source LR systems the only information about the generative model for evidence given the specific source comes from measurements on $o_r$, which are sampled from $r$. Therefore a relatively large number of reference objects (and therefore a large number of measurements) are needed for the parameters of the model for the specific source. Since $r$ varies over cases, this needs to be redone in each case. In contrast, for common-source LR systems, $o_{a1}$ also gives information about the probability of evidence given $H_{1c}$. For common-source LR systems the object collections $o_{a1}$ and $o_{a2}$ are sampled from C, D or T, and can be reused for a multitude of cases, namely for all cases for which C, D or T are assumed representative. This makes common-source LR systems more practical than specific-source LR systems, since once a background database of measurements on $o_{a1}$ and $o_{a2}$ has been built up, they require less work than specific-source LR systems.

---

[7] Sometimes one gets data for 'free'. For example: text-message data from mobile phones from suspects.
[8] This rule of thumb is based on the relations that for $k \geq 1$, $P(LR|H_2 > k) \leq 1/k$ and $P(LR|H_1 < 1/k) \leq 1/k$. Therefore, to accurately model the density of scores under $H_2$ for which the LR = 1000, using $P(LR|H_2 > k) \leq 1/k$ at least 1000 scores under $H_2$ are needed. In analogy, using $P(LR|H_1 < 1/k) \leq 1/k$, it follows that to accurately model the density of scores under $H_1$ for which of LR = 1/100, at least 100 scores under $H_1$ are needed.

Below, the experimental effort for the several types of LR systems is discussed in more detail.

### 3.2.2. Specific-source feature-based LR systems

For specific-source LR systems all information about $r$ comes from $y$. In order to meaningfully derive a LR, parameter uncertainty regarding $r$ should be reasonably small. For example, in the study of Neumann et al. (Neumann and Ausdemore, 2020) it is assumed that the parameters $r$ of the specific source are, in fact, known, so there is no uncertainty about $r$. Now the question is: what experiments are needed to meet the criterion that $r$ can reasonably be assumed known?

Here it is helpful to distinguish two different classes of measurement models for $f(Y|r)$. The first is with negligible randomness in measurement uncertainty, where the outcome of a measurement can be equated to knowing the parameter values $r$. Think of, for example, DNA, where a measured DNA profile on a specific suspect can (with near certainty, ignoring errors like sample switching) be equated to the DNA-profile of this suspect. In that case, one measurement $Y$ suffices to know $r$.

The other option is that there is randomness in $O_r$ (the objects that can be sampled from $r$) and/or in the measurement model. In that case, $r$ can only be known within good precision for a large collection $o_r^i$ and measurements $Y^i$ on $o_r^i$.

This is especially prominent when *Y* comprises a number of features > 1, which is usually the situation. In that situation, one needs a large number of measurements $Y$ for each new case (since each case has a new specific source). This is extremely impractical. Therefore, from a practical standpoint an SSFLR system is virtually impossible to pursue when there is randomness in the model for $P(Y|H_{1s})$.

### 3.2.3. Common-source feature-based LR systems

Common-source feature-based LR systems require only a few measurements on $o_r$. They also use measurements on $o_{a1}$ to estimate $P(X,Y|H_{1c})$, and they use measurements on $o_{a1}$ and $o_{a2}$ to estimate $P(X,Y|H_{2c})$. Often, covariance matrices for multidimensional measurements given $R$ are assumed equal for $r \in R$, and measurements on $o_{a1}$ and $o_{a2}$ are used to estimate the covariance of measurements on $o_r$ (Aitken et al., 2007; Aitken and Lucy, 2004; Aitken and Taroni, 2004; Bolck et al., 2009; Zadora et al., 2013). In that case, relatively few additional measurements on $o_r$ are needed.

For all cases for which the same populations C, D and T are assumed representative, the same measurements on $o_{a1}$ and $o_{a2}$ can be reused.

In general, the more features that are considered, the larger the number of measurements has to be. When only a one-dimensional feature-vector is considered, roughly 3 repetitions of X and Y, per case, and a single collection of measurements on roughly 20 background references each from *C, D* or *T* are needed. For high-dimensional data experimental demands quickly become very challenging and dimension reduction procedures should be applied, see e.g. (Aitken et al., 2019; Vergeer et al., 2020b).

### 3.2.4. Specific-source score-based LR systems

**Non-anchored**

Data for this LR system are partially case-dependent and can partially be reused over cases. It requires, typically, per case a training database of roughly 100 scores under $H_1$ and 1000 scores under $H_2$ (to justify LRs between 1/100 and 1000). In order to accomplish this, for each case, $o_r^i$ and $o_t^i$ need to be sampled from $r$ and measured (and typically $i = 1, ..., 100$, in order to get 100 conditionally independent scores under $H_1$). Apart from this measurements on different traces from the alternative background population of traces (*T*) need to be generated (typically 20, which can each be compared to the 100 measurements on the reference, which gives 2000 scores under $H_2$, but they are not identically and independently distributed (i.i.d.)). The 20 measurements on *T* can be re-used over cases.

It can be a daunting task to generate the required amount of 100 scores for the distribution under $H_1$ for each new case. However, there may be a shortcut. If, given $r$, the distributions of $O_r$ and $O_t$ are identical, an ad-hoc approach (violating the i.i.d. assumption however) is to generate for each case, say, 20 objects from the specific source (and one measurement each), and cross-compare each. In this way 20 * 19 /2 = 190 comparisons are obtained given $H_1$. When, for the distribution under $H_2$, the 20 measurements are each compared to, say, 70 measurements from the alternative background population of traces (*T*), 1400 comparisons under $H_2$ are generated (although not i.i.d.). These are approximately the required number of scores needed to justify LRs between 1/100 and 1000.

What seems strange is that for the case at hand, measurements $y^i$ with $i = 1, ..., n$ are used to estimate the required score distributions while one more measurement $y^{n+1}$ is used for the score for which an LR is evaluated. Which measurement is used for which purpose seems arbitrary.

**y-anchored**

For a y-anchored SSSLR $y$ is fixed, as well as the reference itself. It requires, typically, per case measurements on 100 objects from $o_r{}^i$ (one measurement $x^i$ each, and assuming there is variation in $O_r$) generated from $r$. These can each be compared to $y$ to get 100 scores $\delta$ given $H_{1s}$ and $y$. For the score-distribution under $H_{2s}$, typically 1000 measurements $x^i$ (so $i = 1, ..., 1000$) following all sampling steps starting from the alternative background population of traces ($T$) are needed, which should each be compared to $y$. The latter measurements can be reused over cases. The 1000 $H_{2s}$ and the 100 $H_{1s}$ combinations justify LRs between 1/100 and 1000.

It seems to be an extremely challenging task (for most evidence modalities in forensic science) to generate the required number of 100 scores for the distribution under $H_{1s}$ for each new case.

### 3.2.5. Common-source score-based LR systems

**Non-anchored**

Common-source non-anchored SLR systems require measurements on two background populations (C, and D or T) to estimate the score-distributions under $H_{1c}$ and $H_{2c}$. These measurements can be used repeatedly over cases. This makes this type of approach very practical. Typically, measurements (one each) on 200 objects $o_{a1}{}^i$ and 20 objects $o_{a2}{}^j$ are needed. Combining $o_{a1}{}^i$ pairwise (and calculating a score per pair), gives 100 i.i.d. scores under $H_1$. Calculating a score for each object in $o_{a1}{}^i$ to each object $o_{a2}{}^j$ gives 4000 scores under $H_{2c}$ (although not i.i.d.). All these measurements can be reused over cases. Furthermore, in each case a measurement on the trace and reference in the case is required to get the score in the case.

**y-anchored**

One still needs measurements on the two background populations *C* and *D* or *T*, which can be re-used over cases. However, in order to estimate $f(\delta(x,y)|H_{1c}, y)$ using the measurements on *C* one needs to select those sources where the measurement on $o_r{}^i$ gave a result close to $y$ (see the path analysis in Section 2.4.2.2). Therefore, one needs a multitude of measurements on *C* when compared to the non-anchored counterpart. Now let us suppose the number of measured variables is one. In order to distinguish, say, 50 categories of *y* to condition on, one needs under $H_{1c}$ typically measurements on 5000 sources from *C* (100 * 50) and also 5000 measurements on traces generated from these sources, in order to select 100 relevant pairs given $H_{1c}$ and $y$.

Also, for the denominator of the LR, when *C* is different from *D*, one needs 5000 sources from D as well (in order to select 100 based on measurement result close *y*). From *T* one needs to select 20 sources (one measurement each) to get 2000 scores under $H_{2c}$ when *y* is fixed. In conclusion, for most evidence modalities to measure so many objects from *C* and *D* is very challenging. Luckily, measurements from *C*, *D* and *T* can be reused over cases. Finally, one needs a measurement of *x* and *y* for the trace and reference in the case at hand.

**x-anchored**

Experimental requirements are the same as for the y-anchored system except that one now needs to select, based on *x*, a subsection of measurements on sources from the population *T* under $H_2$.

### 3.3. Performance ranking versus practical feasibility

Figure 2 orders the classes of LR systems on performance (Y-axis) and experimental demands (X-axis). In Figure 2 performance is framed in terms of 'information loss', using the information theoretical interpretation of the logarithmic SPSR. Performance loss (and information loss) can be restricted to certain dimensions (in this case X, Y, R, or a combination), which is shown in the plot. An LR system may lose information in a certain dimension, for example when the likelihood function is defined as an average over this dimension. Information loss can be regarded as the negative of performance; the larger the information loss, the worse the performance.

From Figure 2 it becomes clear that roughly there is a trade-off in information loss and experimental demand. Where the SSFLR system has no information loss, it is unfortunately virtually impossible from a practical point of view for the commonly encountered situation where there is uncertainty concerning a measurement Y given $r$, and the number of features exceeds 1. On the other extreme CSSLR systems have the least experimental demand but the most information loss. In between are the other classes of LR systems.

CSFLRs come out favourably in terms of performance and experimental demands. While the information loss is restricted to the R-dimension, its experimental demands require a one-time investment for repeated use in casework. This one-time investment grows when the dimension of the feature-vector increases. The use of dimension-reduction techniques can keep experimental demands within reach.

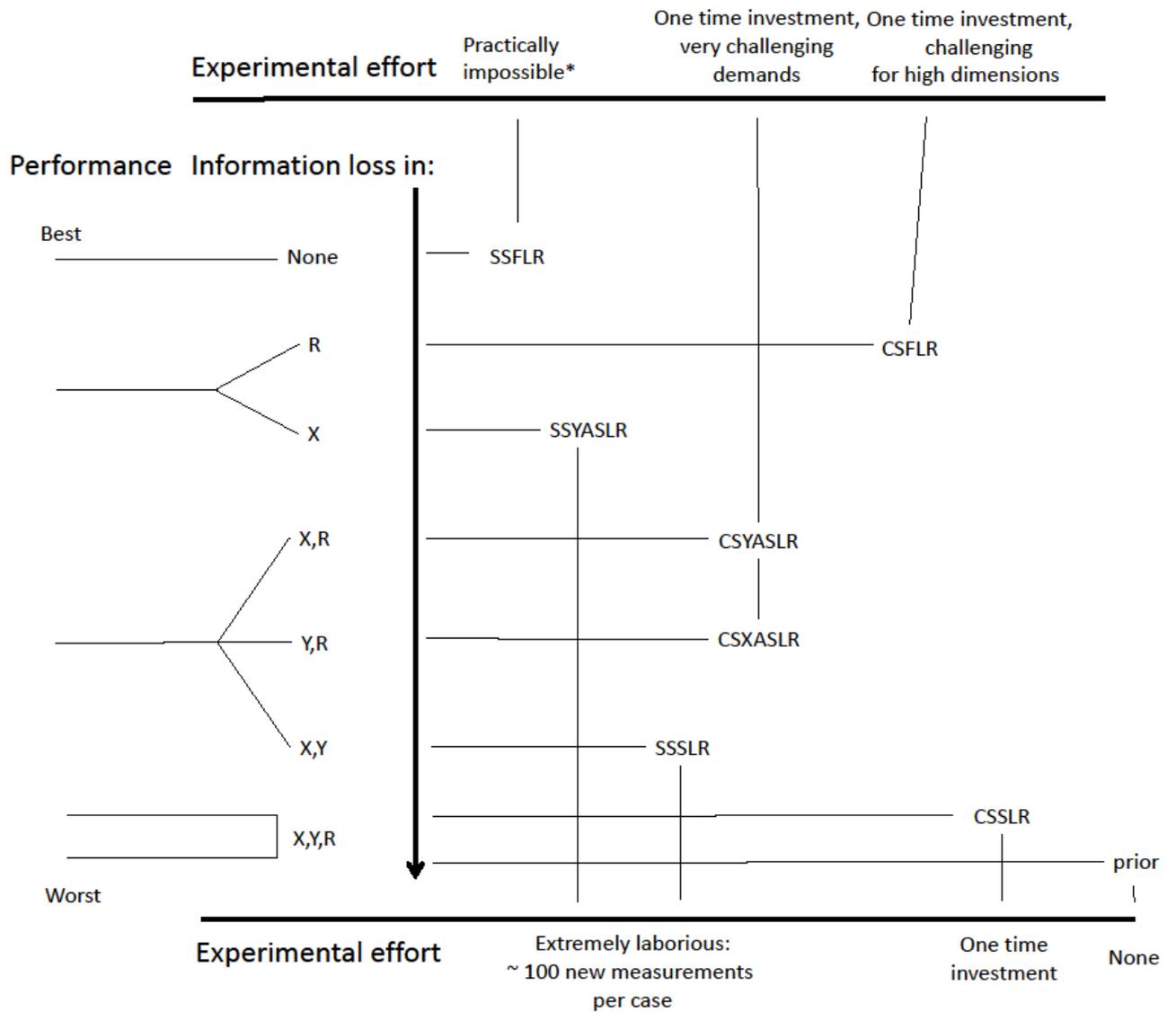

Figure 2. Overview of performance and experimental effort for the classes of LR systems. The scale for both axes is ordinal (for x-axis the upper and lower axes are on the same scale).

# 4. Discussion and conclusion

## 4.1. Summary of results

In this paper, different classes of source-level LR systems were studied for the scenario of comparing a trace to an identified reference. The most important results are as follows, if LR systems are well-calibrated:

1. Common-source LRs go well with trace-reference priors, see Section 3.1.4.
2. Score-based LR systems without rarity can be used, and perform better than prior only, see Section 3.1.3.
3. LR systems that are preferable from a performance point of view are often hard to realise from a practical point of view, see Figure 2.
4. A trade-off between performance and experimental demand is observed, with the one positive exception a CSFLR system, see Figure 2.

In Sections 4.2 to 4.9 several topics are discussed.

## 4.2. Position in the 'true LR' debate

The analysis in this paper concurs with the statement that 'the LR does not exist' (Berger and Slooten, 2016). There is no 'true' or 'benchmark' value of the likelihood ratio, its value depends on training data, modelling assumptions and what is used as evidence. The evidence is different when comparing score-based LR systems (where indeed the evidence is the score) to feature-based LR systems (where the evidence are the features). Naturally, different values arise for a score-based LR and a feature-based LR. Neither one of these single LR values is regarded as 'true' or 'wrong' o r 'better' based on some objective standard. However, performance of a sequence of LRs for which the ground truth is known can be studied by using SPSRs, comparing LR systems by average performance. On average, feature-based LR systems outperform score-based LR systems.

## 4.3. Is an LR system suitable? Comparing LR systems to each other or comparing them to 'prior only'.

Regarding point 2 in the summary of results: in the literature there has been some debate about the supposedly unsuitability of similarity-only scores (Morrison and Enzinger, 2018; Neumann et al., 2020; Neumann and Ausdemore, 2020). These studies had a different perspective than the present study to substantiate their argument. Morrison and Enzinger (Morrison and Enzinger, 2018) take a feature-based LR value as a benchmark value and then show that the score-based LR value is different, and moreover that SLR values are the same for references with different *r* but the same score, since a measure of rarity was not included in the score.

Neumann, Hendriks and Ausdemore ( 2020) use an analogous kind of setup in comparing LR systems: specific-source feature-based LRs are compared to common-source feature- or score-based LRs taking the specific-source feature-based LRs as the benchmark, and generating a number of LRs for a specific trace or reference. One of their arguments to discard score-based LR systems is that they notice 'a lack equality (sic) between the Bayes factors'. Furthermore, they state 'none of the score-based models proposed to date can be considered as suitable proxies of the Bayes factor of interest', and their 'Bayes factor of interest' is the feature-based specific-source likelihood ratio.

I agree that indeed LR values will differ in these experimental setups, and scores can disregard rarity. However, the fact that LR values are different does not mean that one of the two LR systems is wrong. They both reflect evidential value, but for one a different statistical model (i.e. different class of LR system) is used than for the other. Disregarding rarity in a score means that expected performance for such a model may be less than for a feature-based model.

However, both classes of LR systems may be suitable in the sense that expected performance over cases (in terms of the resulting posterior odds) is better than just using prior odds, as shown in this paper.

When designing an LR system theoretical optimal performance is not the only criterion however. One also has to consider whether one is able to find a well-fitting model for the features, and how demanding the experimental burden is to get reasonable parameter estimates for these models. As shown in this paper, experimental burden varies strongly for the different classes of LR systems.

### 4.4. Possible criticism: 'some LR systems are biased against certain suspects'

Some forensic scientists may state that some LR systems are biased against certain suspects, while others are not. Neumann and Ausdemore (2020) and Neumann, Hendricks, and Ausdemore (2020) show that common-source SLRs may result in a larger expected LR value (when compared to specific-source feature based LRs) for common suspects when $H_2$ is true. Clearly, this is an unwanted bias.

Unfortunately, it is extremely hard to design an unbiased LR system. The criticism above suggests that specific-source feature-based LR systems are unbiased. I argue that, unfortunately, this is not true for most evidence types, due to a variety of reasons. First, there is often a selection of included features (which data from a finger print, which chemical elements, which loci, etc.). Excluding certain features (either by not measuring them or not selecting them) will certainly be disadvantageous to certain suspects. Second, in order to reduce experimental burden, dimension reduction techniques are often applied. Again, this will be disadvantageous to certain suspects. Third, for feature-based LR systems statistical model families are assumed that often do not comply with the unknown data generative models. This, again, introduces bias against certain suspects. Unfortunately, at the moment it is very hard (if not impossible) to avoid these biases.

The difference between these sources of bias is that for score-based LR systems they may be apparent (looking at the raw measurements, one may infer that one is dealing with a common suspect), while for feature-based LR systems they may not be apparent (looking at the raw measurements, one does not know whether using the feature selection/dimension reduction/misfitted model results in bias against this suspect). For some, this may be a reason to disqualify score-based LR systems. However, one may also defend the opposite and disqualify specific-source feature-based LR systems for this reason: when a possible bias can be identified (as may be the situation for score-based LR systems), one is in principle able to take measures to correct for it. When a possible bias cannot be identified (as is the case for feature-based LR systems), one can never correct for it.

However, I think neither LR systems should be disqualified: at the moment one inevitably has to deal with LR systems that may be biased against certain suspects. We will have to live with the observation that - at the moment - the best we can do is to design an LR system that performs well on average, over a number of suspects (i.e.: court cases). In the advantageous situation when bias against a certain suspect can be identified, one may adjust the LR value accordingly. However, in that situation one usually has to rely on expert opinion for the adjustment, since if this could have been done automatically there would be no reason to not include it in the LR system already.

### 4.5. SSFLR systems with or without epistemic uncertainty?

The current paper has studied SSFLR systems without epistemic uncertainty about the sampling model for reference evidence, i.e. $r$ is known or there are enough measurements in $Y$ to model the evidence when the specific source is involved. In Section 3.2.2 of the present paper it was remarked that, when $Y$ is multidimensional and there is aleatoric uncertainty in the object parameter values or measurement uncertainty, complete knowledge of this model requires a very large number of repeated measurements $Y$ on the specific reference.

However, in the literature SSFLRs with epistemic uncertainty have been introduced by Ommen et al. (Ommen et al., 2017); they require less measurements on the specific source. These authors do not assume $r$ is known. Information about $r$ is obtained through a limited number of measurements $Y$ from $o_r$. In order to calculate the LR, as Bayesian calculus prescribes, they postulate a prior distribution over parameter space $R$ and integrate over it (approximately, by using Markov chain Monte Carlo methods). The result is an SSFLR 'averaged over $R$', also called a full-Bayesian LR or Bayes factor. This is a solution to the problem that for specific-source feature-based LRs a lot of measurements $Y$ are needed, and it adheres to the rules for Bayesian probability calculus postulating a probability distribution for unknown parameter values.

This compares well with the CSFLR as defined in the present paper, since here $R$ is also integrated over. Interestingly the mathematical description in terms of integrals over $R$ is the same for the two approaches, only the choice of prior distribution for $R$ is different. Where the common source LRs have an empirically motivated prior distribution (the specific source, and therefore $r$, is a random sample from a population of relevant references, sampled by $o_a$), Ommen et al. provide no such motivation for the prior choice for $r$ for their specific source Bayes factor.

From this perspective, specific-source feature-based Bayes factors may possibly have less performance (as judged by strictly proper scoring rules) than CSFLRs, since for the latter the prior distribution for $r$ relies on real-world information, present in $o_a$. This is an interesting topic for further study.

### 4.6. Which evidence?

The relationships between the performances of the several classes of LR systems discussed above hold under the presumption that the same features are used for feature-based LR systems or for calculating scores based on these features for score-based LR systems. In practice however, one may have the choice to either develop a feature-based LR system based on a limited number of features (since this is practically more feasible than for all the features) or develop a score-based LR system based on all the features. This for example comes about for data based on elemental composition, where the concentrations of numerous elements are measured. In the literature (Corzo et al., 2018; van Es et al., 2017; Vergeer et al., 2020b) feature-based LR systems are implemented for a subset of the measured features (in combination with a post-hoc calibrating step, see below), while a score-based LR system for all the features may outperform these feature-based LR systems.

### 4.7. The importance of being well-calibrated

Note that the results presented above are only valid under the condition that the LR systems output well-calibrated LRs, meaning that assumed distributions comply with real-world data generating processes. When LR systems are ill-calibrated, even when one uses the in principle best performing SSFLR system, it is possible to perform worse (in terms of posterior odds) than prior-only. This underlines the importance of developing well-calibrated LR systems.

Score-based LR systems are much more easy to calibrate than feature-based LR systems, since when the one-dimensional score-distributions are modelled by appropriate density functions (for example, using Kernel Density Approximation (Silverman, 1998)), the resulting LRs are usually well-calibrated. On the other hand, ill-calibrated LRs from feature-based LR systems may be calibrated post-hoc by calculating feature-based likelihood ratios for a test set under $H_1$ and $H_2$ and subsequently calibrating this likelihood-ratio output (which is one-dimensional) by the same methods as one calibrates scores (Morrison et al., 2021; Vergeer et al., 2020a).

Note however, that despite common usage of post-hoc calibrating in the field of speaker recognition (Morrison et al., 2021) and its introduction to other forensic science disciplines (Corzo et al., 2018; van Es et al., 2017; Vergeer et al., 2020b), there is scientific debate stating demerits (Aitken et al., 2020, 2019) or merits (Vergeer et al., 2020a) of a post-hoc calibrating step.

### 4.8. Replies to criticisms against certain classes of LR systems

The results and discussion paragraphs above can be used to formulate a reply to the three criticisms on certain classes of LR systems mentioned in the introduction:

i. *Common-source LR systems are irrelevant to questions in court when a suspect (or reference object) is identified (Ommen and Saunders, 2018).* This paper shows that when averaging over court-cases, posteriors using common-source LR sampling models do have merit over just using the trace-reference prior probability.

ii. *Some classes of LR systems may unpredictably over or underestimate the weight of the evidence represented by a trace and a reference object (Neumann and Ausdemore, 2020).* In practice LR systems are meant for repeated use in casework. In a case, one has the choice to either use an LR system or not. Therefore, it is more relevant to compare, averaged over cases, the use of an LR system (for updating the prior odds) to not using the LR system and using prior odds only. From this perspective, all LR systems have merit. However, some LR systems have better expected performance (in terms of posterior odds) than others. On the other hand, practical considerations are also important, since some classes of LR systems have harsh experimental demands.

   iii. *Score-based LR systems that do not include a measure of rarity do not qualify as evidence appropriate to compute likelihood ratios (Morrison and Enzinger, 2018).* Score-based LR systems without rarity can be used, on average posteriors using these LR systems perform better than prior only.

### 4.9. Instead of a 'common-source' and a 'specific-source' problem, shouldn't we speak of a trace-trace and a trace-reference problem?

In light of the results in this paper, we can distinguish between two different types of origin-of-source problems and two different types of models to address these problems. The distinction is:

- A trace-trace problem, which should always be addressed with a common-source model, using samples from background populations of traces.
- A trace-reference problem for which one is unable to obtain a sample from a relevant background population of references. It should always be addressed by a specific-source model.
- A trace-reference problem for which one is able to obtain a sample from a relevant background population of references. It may be addressed by a specific-source model (and then one does not use this background population for information about parameters describing the reference) or by a common-source model (and then one uses the background population of references for information about parameters describing the reference).

For the latter type of problem, it is also a matter of experimental convenience which class of LR system one prefers, as explained in Section 3.2.1.

Following the above distinction, I think it is wise not to speak of common or specific-source problems or hypotheses anymore, since I think it is confusing language: common-source sampling models can be applied to trace-reference (or in present literature 'specific-source') problems. Therefore, I think it is better to speak of a trace-trace or a trace-reference origin of source problem. And when addressing a trace-reference problem, to mention the sampling model used to model evidence where the reference is involved. This can be specific-source: only $Y$ is used to model $r$, or this can be common-source: $r$ is also modelled by using a random selection of references from a relevant background population.

## 4.10. Conclusion

This paper shows that LR systems based on common-source models can be used for the trace-reference problem. For trace-reference origin of source problems, it unifies common- and specific-source LR systems, from feature- to score-based, in an overall ranking system, using strictly proper scoring rules. Furthermore, it discusses practical demands to realise the systems. This overview helps forensic scientists to better understand and more deliberately choose between the several classes when designing LR systems for trace-reference problems, and it is also meant to create clarity within the scientific discussion on the supposed unsuitableness of some classes of LR systems to the trace-reference problem. This paper has 'ranked the stars', and an important outcome is that, apart from specific-source x-anchored LR systems, all well-calibrated and well-conditioned LR systems have merit over the prior odds. I hope that this helps forensic science to progress and forensic scientists to reach for the stars when designing LR systems, since, from the perspective of evaluating LR systems by strictly proper scoring rules, we are a lot more flexible in design demands than many may have realised.

## Acknowledgements

I am sincerely grateful to my colleagues Elina Sergidou, Kim de Bie, Leen van der Ham, Marjan Sjerps, Timo Matzen and Wauter Bosma for proof reading an earlier version of the manuscript and helping me to redirect and improve readability. Also, I am eternally grateful to Bart Bruins for his invaluable moral support.

## Funding

This research received no external funding.

# Appendix 1. Path analyses for the different classes of LR systems

$$SSFLR := \frac{f(x,y|H_{1s})}{f(x,y|H_{2s})},$$

Numerator: sample $o_t^i$ and $o_r^i$ from $r$. I use the index '$i$' here to denote that $o_t^i$ and $o_r^i$ have one of many possible 'values', one of these possible values is $o_t$ and $o_r$. After obtaining $o_t^i$ and $o_r^i$ by sampling from $r$, next sample $x^i$ and $y^i$ from $o_t^i$ and $o_r^i$ respectively. Do all the sampling steps over and over again. Count the relative frequency that one sees $\{x,y\}$. The denominator can be evaluated as follows: sample $t^i$ from T, sample $o_t^i$ from $t^i$ and sample $o_r^i$ from $r$, sample $x^i$ and $y^i$ from $o_t^i$ and $o_r^i$ respectively. Do all the sampling steps over and over again. Count the relative frequency that one sees $\{x,y\}$.

$$CSFLR := \frac{f(x,y|H_{1c})}{f(x,y|H_{2c})}$$

Numerator: sample $r^i$ from C, sample $o_t^i$ and $o_r^i$ from $r^i$, sample $x^i$ and $y^i$ from $o_t^i$ and $o_r^i$ respectively. Do all the sampling steps over and over again. Count the relative frequency that one sees $\{x,y\}$. Denominator: sample $t^i$ from T, and $r^i$ from D, sample $o_t^i$ from $t^i$ and $o_r^i$ from $r^i$, sample $x^i$ and $y^i$ from $o_t^i$ and $o_r^i$ respectively. Do all the sampling steps over and over again. Count the relative frequency that one sees $\{x,y\}$.

$$SSSLR := \frac{f(\delta(x,y)|H_{1s})}{f(\delta(x,y)|H_{2s})}$$

Numerator: sample $o_t^i$ and $o_r^i$ from $r$, sample measurements $x^i$ and $y^i$ from $o_t^i$ and $o_r^i$ respectively. Calculate $\delta^i$. Do all the sampling steps over and over again. Count the relative frequency that one sees $\delta$. Denominator: sample $t^i$ from T, sample $o_t^i$ from $t^i$ and $o_r^i$ from $r$, sample $x^i$ and $y^i$ from $o_t^i$ and $o_r^i$ respectively. Calculate $\delta^i$. Do all the sampling steps over and over again. Count the relative frequency that one sees $\delta$.

$$CSSLR := \frac{f(\delta(x,y)|H_{1c})}{f(\delta(x,y)|H_{2c})}.$$

Numerator: sample $r^i$ from C, sample $o_t^i$ and $o_r^i$ from $r^i$, sample measurements $x^i$ and $y^i$ from $o_t^i$ and $o_r^i$ respectively. Calculate $\delta^i$. Do all the sampling steps over and over again. Count the relative frequency that one sees $\delta$. Denominator: sample $t^i$ from T and $r^i$ from D, sample $o_t^i$ from $t^i$ and $o_r^i$ from $r^i$, sample $x^i$ and $y^i$ from $o_t^i$ and $o_r^i$ respectively. Calculate $\delta^i$. Do all the sampling steps over and over again. Count the relative frequency that one sees $\delta$.

$$SSYASLR := \frac{f(\delta(x,y)|H_{1s},y)}{f(\delta(x,y)|H_{2s},y)},$$

Numerator: sample $o_t^i$ from $r$, sample measurements $x^i$ from $o_t^i$. Calculate $\delta^i(x^i,y)$. Do all the sampling steps over and over again. Count the relative frequency that one sees $\delta$. Denominator: sample $t^i$ from T, $o_t^i$ from $t^i$, sample measurements $x^i$ from $o_t^i$. Calculate $\delta^i(x^i,y)$. Do all the sampling steps over and over again. Count the relative frequency that one sees $\delta$.

$$CSYASLR := \frac{f(\delta(x,y)|H_{1c},y)}{f(\delta(x,y)|H_{2c},y)}.$$

Numerator: sample $r^i$ from C, sample $o_t^i$ and $o_r^i$ from $r^i$, sample measurements $x^i$ and $y^i$ from $o_t^i$ and $o_r^i$ respectively. Keep all paths for which $y^i$ equals $y$. Calculate $\delta^i(x^i,y)$. Do all the sampling steps over and over again. Count the relative frequency that one sees $\delta$ in the kept paths. Denominator: sample $t^i$ from T, sample $o_t^i$ from $t^i$, sample measurements $x^i$ from $o_t^i$. Calculate $\delta^i(x^i,y)$. Do all the sampling steps over and over again. Count the relative frequency that one sees $\delta$.

$$SSXASLR := \frac{f(\delta(x,y)|H_{1s},x)}{f(\delta(x,y)|H_{2s},x)},$$

Numerator: sample $o_r^i$ from $r$, sample measurements $y^i$ from $o_r^i$. Calculate $\delta^i(x, y^i)$. Do all the sampling steps over and over again. Count the relative frequency that one sees $\delta$ in the paths.

Denominator: sample $o_r^i$ from $r$, sample measurements $y^i$ from $o_r^i$. Calculate $\delta^i(x, y^i)$. Do all the sampling steps over and over again. Count the relative frequency that one sees $\delta$ in the paths. The two methods for path generation are identical, so the LR = 1.

$$CSXASLR \coloneqq \frac{f(\delta(x,y)|H_{1c}, x)}{f(\delta(x,y)|H_{2c}, x)}.$$

Numerator: sample $r^i$ from C, sample $o_t^i$ and $o_r^i$ from $r^i$, sample measurements $x^i$ and $y^i$ from $o_t^i$ and $o_r^i$. Keep all paths for which $x^i$ equals $x$. Calculate $\delta^i(x, y^i)$. Do all the sampling steps over and over again. Count the relative frequency that one sees $\delta$ in the kept paths. Denominator: sample $r^i$ from D. Sample $o_r^i$ from $r^i$, next sample $y^i$ from $o_r^i$. Calculate $\delta^i(x, y^i)$. Do all the sampling steps over and over again. Count the relative frequency that one sees $\delta$ in the paths.

# Appendix 2: A cost-benefit analysis of LR systems

The section below does not contain any information that is scientifically new. However, some readers may be unfamiliar to strictly proper scoring rules, and this introduction is written in order to fully appreciate their merits for evaluating performance of LR systems.

## A.1. Evaluate likelihood ratios or probabilities?

Ultimately, the judicial system is not interested in likelihood ratios, but in decisions based on posterior probabilities. The benefit of LR systems may accordingly be measured in how they improve posterior probability distributions compared to prior probability distributions. This improvement can be evaluated by strictly proper scoring rules. Therefore the paper is focused on evaluating the performance of (posterior) probability distributions, by strictly proper scoring rules (SPSRs).

## A.2. What makes strictly proper scoring rules good scoring rules?

A scoring rule is a rule for scoring probability statements depending on the magnitude of the stated probabilities (i.e. the stated probability distribution over possible events) and the associated truth (the event that has occurred). So the question is: what would make up a good scoring rule?

### A.2.1. A scoring rule should promote discriminating probability statements

First, one would like to reward discriminating probability statements (probability statements that are closer to 0 or 1). In order to promote discriminating probability statements, one should use as a scoring rule any function for which the expected value increases when one states a probability closer to the truth. In technical terms: the scoring rule should promote discrimination (Degroot and Fienberg, 1983).

### A.2.2. A scoring rule should promote honest probability statements

Second, the scoring rule should promote 'honesty', i.e. forecasters should be encouraged to state probabilities that they actually believe to reflect accurately their state of knowledge.

### A.2.3. Strictly proper scoring rules meet both criteria

Not any scoring rule that increases its score when the stated probabilities become more discriminating promotes honesty. For example, a scoring rule that would meet the discrimination criterion but not the 'honesty' criterion is the scoring rule in Table 3. For simplicity, probabilities are made categorical with steps of 0.1. Let's say that one wants to predict the event that 'it will rain some time tomorrow'. It shows the score as a function of P(rain), when it did rain (second column) and when it did not rain (third column). This scoring rule promotes discrimination: the expected score increases when probability statements are closer to 0 or 1 (which is shown in SPSRs are defined as scoring rules that meet both the discrimination and the honesty criterion (Degroot and Fienberg, 1983). Examples of SPSRs are the logarithmic scoring rule or the quadratic (or Brier) scoring rule (Brümmer, 2010; Brümmer and Du Preez, 2006). There are infinitely many SPSRs[9], each having a unique scoring function.

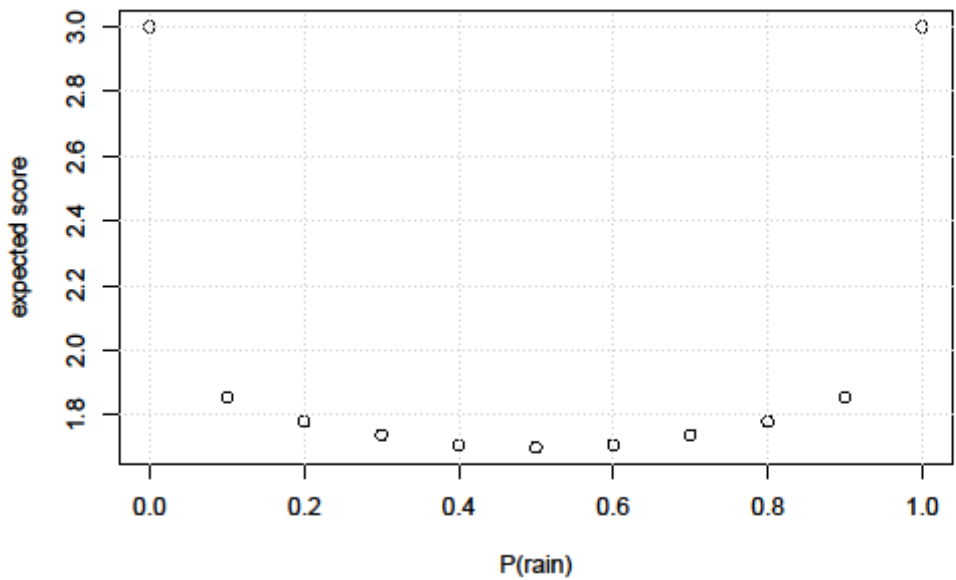

Figure 3), but so strongly for P(rain) = 0 or 1 that the forecasters are inclined to be dishonest in order to optimise their expected score. Indeed, when they believe their P(rain) to be 0.1 or 0.9, they would be better off stating that their P(rain) = 0 or 1 respectively. The expected score for stating "P(rain) = 0.1" is $0.1 \times 1 + 0.9 \times 1.95 = 1.86$, while the expected score for stating a dishonest "P(rain) = 0", while they actually believe it to be "0.1" is $0.1 \times 0 + 0.9 \times 3 = 2.7$, which is a larger expected score. Therefore, this scoring rule promotes dishonesty.

| P(rain) | Score when it rained | Score when it did not rain |
|---|---|---|
| **0** | 0 | 3 |
| **0,1** | 1,00 | 1,95 |
| **0,2** | 1,30 | 1,90 |
| **0,3** | 1,48 | 1,85 |
| **0,4** | 1,60 | 1,78 |
| **0,5** | 1,70 | 1,70 |
| **0,6** | 1,78 | 1,60 |
| **0,7** | 1,85 | 1,48 |
| **0,8** | 1,90 | 1,30 |
| **0,9** | 1,95 | 1,00 |
| **1** | 3 | 0 |

Table 3. Score as a function of stated probability P of raining the next day, when it did or when it did not.

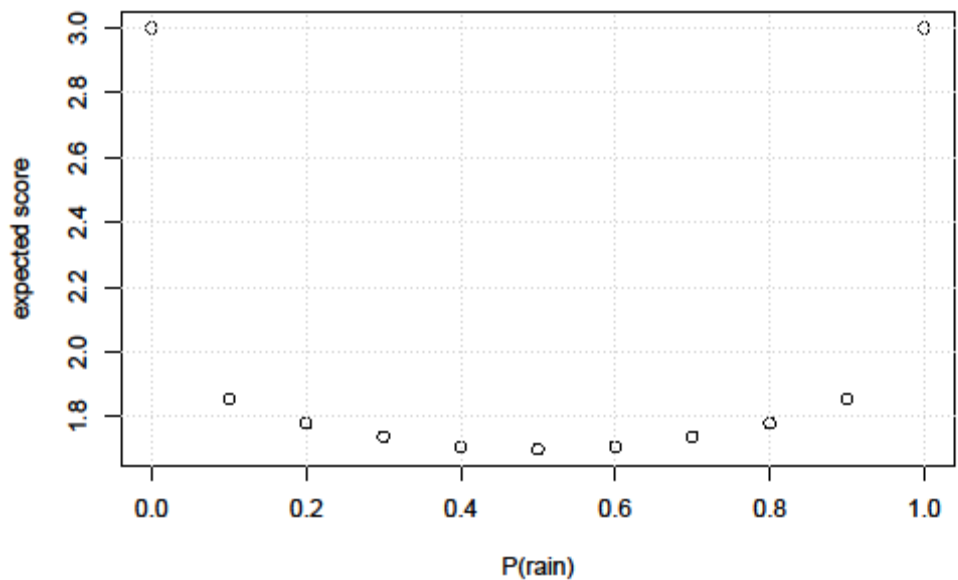

Figure 3. Expected score as a function of P(rain) for the scoring function in Table 3.

Formally a SPSR is a function of the stated probability distribution S(Y) over the event space Y and the observed outcome (which is one of the events), y:

$$\text{SPSR} := f[S(Y), y]$$

The expected score is then defined as the expected value with regard to y. This value does not only depend on the stated probability distribution $S(Y)$, but also on what probabilities one actually believes to be the accurate probabilities for the outcomes. Let $B(Y = y)$ be the probability distribution on the same event space $Y$ that one believes to accurately reflects ones knowledge about Y. Then:

$$E_Y(f[S(Y), y]) = \sum_y B(Y = y) f[S(Y), y], \qquad \text{(Eq. A1)}$$

and as stated above the expected score is maximised when $S(Y) = B(Y)$ (expected score is maximal when one reports 'honest' probability statements).

Therefore, by definition of using an SPSR as evaluation function the expected score stating $B(Y)$ (ones belief, or a known true probability distribution) cannot be smaller than the expected score stating $S(Y)$ (Degroot and Fienberg, 1983; Gneiting and Raftery, 2007):

$$\sum_y B(Y = y) f[B(Y), y] \geq \sum_y B(Y = y) f[S(Y), y] \qquad \text{(Eq. A2)}$$

---

[9] A strictly proper scoring rule can be based on any function which is convex in the domain [0,1].

*A.2.4. Relation of SPSRs to decisions made in court*

Evaluating probability distributions by SPSRs has a connection to evaluating decisions in court, under 2 assumptions. Firstly: decisions in court are based on a posterior probability of guilt, and secondly: the decision is made according to some (unknown) threshold probability. If the posterior probability of guilt is larger than the threshold value, the suspect is found guilty. If the posterior probability of guilt is smaller than the threshold value, the suspect is found not guilty.

If for court cases this threshold value was known, one could derive a scoring rule using this threshold value (Biedermann et al., 2008; Lindley, 1991) and it would most certainly be more interesting to evaluate probability distributions against this case-specific scoring rule than against any strictly proper scoring rule. However, they are not known, and moreover it seems reasonable to assume that thresholds vary over court cases due to a difference in circumstances unrelated to the probability analyses. They may also vary among judges and juries.

Therefore, at the moment, it is best to evaluate probability distributions against a distribution of thresholds. Interestingly, each specific SPSR is related to a specific distribution of thresholds (Brümmer and Du Preez, 2006). And since there are infinitely many SPSRs, a great variety of distributions of thresholds is covered by studying results that hold for all SPSRs, which is done in the present paper.

Even more compelling is that court's decisions between guilty and not guilty may be interpreted as applying a proper[10] scoring rule: decisions of guilt/not guilt based on a comparison of one's 'honest' probability of guilt with a threshold probability results in optimal expected score. The threshold probability value may be determined from the rewards for 'good decisions' (acquitting the innocent and convicting the guilty) and the costs for 'wrong decisions' (convicting the innocent and acquitting the guilty). Probability statements that maximise expected score for strictly proper scoring rules also maximise expected score for proper scoring rules, like the ones for 'hard decisions' in court cases. Therefore, all results derived in this paper are also valid for a decision in a single court case if it is based on (implicitly) comparing a posterior probability to some threshold value.

---

[10] It is not strictly proper since the solution that leads to the maximum expected score is not unique. For example, when a 'honest' probability of guilt is larger than the threshold, any number representing the probability of guilt results in the same decision, as long as it is larger than the threshold value.

### A.2.5. SPSRs promote well-calibrated probability statements

An additional benefit from evaluation by strictly proper scoring rules is that they promote well-calibrated probabilities (Brümmer and Du Preez, 2006; Degroot and Fienberg, 1983). Measuring calibration of a sequence of probabilities relates (Bayesian) probability statements to relative frequencies; it asks the question: do (Bayesian) probability statements have the same value as their relative frequency analogues? For example, and using a weather forecasting example (Dawid, 1982): from the days for which a weather forecaster stated that the (Bayesian) probability of rain was 0.5, did it actually rain half of the times?

For a dataset of well-calibrated probabilities $P$, the following relation holds (approximately[11]) for the stated probabilities and relative frequencies of these probabilities in the data.

$$rel\ freq(P = x) = x \quad\quad\quad\quad (Eq.\ A3)$$

### A.2.6. Scoring rules should be evaluated in the long run

Assume that one is interested in measuring the performance of a weather forecaster who, for each day, states a probability P(rain) on the chance of rain the next day. And we keep track of whether it actually rained that day. We measure average performance by repeatedly putting a score on the stated probability distribution when compared to the truth (as a strictly proper scoring rule does). The averaging helps to accurately measure performance by 'sampling' from all possible probability distribution statements of the forecaster.

---

[11] This relation is approximate for two reasons. Firstly, counting frequencies requires defining bins while probabilities are on a continuous scale. Secondly, the data are a finite sample from a population.

When scoring only a few instances, the obtained average score has high variance. To reduce the variance and reliably judge the quality of the forecasts, one must calculate the score for a sequence of forecasts, and not for a single case where the forecaster may be lucky or not. It may be argued that the same reasoning holds for probability statements with respect to court cases. LR systems should be evaluated over the sequence of cases that they will be used for, it has very little meaning to draw conclusions about the performance of an LR system based on a single case, person or object. LR systems are meant to be used on a number of cases, persons or objects. In one case one LR system may do better, while in a next case another LR system may do better. Since scoring rules optimise expected score, they are also suitable to measure score in the long run, over a sequence of cases.

### *A.2.7. Related work that uses SPSRs to evaluate LR systems*

The idea to use an SPSR to evaluate performance of LR systems is not new. Ramos et al. (Ramos et al., 2013; Ramos and Gonzalez-Rodriguez, 2013) and Brümmer et al. (Brümmer and Du Preez, 2006) use an empirical version (for a dataset of LRs acquired by experiment) to assess performance of single LR systems or to compare LR systems to each other. The SPSR that is used is the logarithmic SPSR, which is central to information theory (Cover, 2005).

This paper uses SPSRs, not to evaluate or compare single LR systems, but to compare the relative performance of the general classes of LR systems introduced in Section 2.4.

### *A.2.8. Example of a strictly proper scoring rule: the logarithmic scoring rule*

The logarithmic scoring rule is an example of a strictly proper scoring rule. In Brümmer, (2010) page 39, it is explained why:

The scoring rule is: $log_2(Q)$, with $Q$ the stated probability of the event that happened.

This scoring rule promotes honesty: the expected value for a binary event with true probabilities $P(H_1)$ and $P(H_2)$ is:

$$P(H_1)log_2\big(Q(H_1)\big) + P(H_2)log_2\big(Q(H_2)\big)$$

Taking the derivative with respect to $Q(H_1)$ one finds that this function is maximal when $Q = P$. Therefore, expected score is maximised by stating one's true belief.

Moreover, this score promotes discrimination. The expected score is larger for probabilities that deviate from $P = 0.5$. This can for example be seen by plotting $P(H_1)$ versus $P(H_1)log\big(P(H_1)\big) + P(H_2)log\big(P(H_2)\big)$, which is shown in Figure 4.

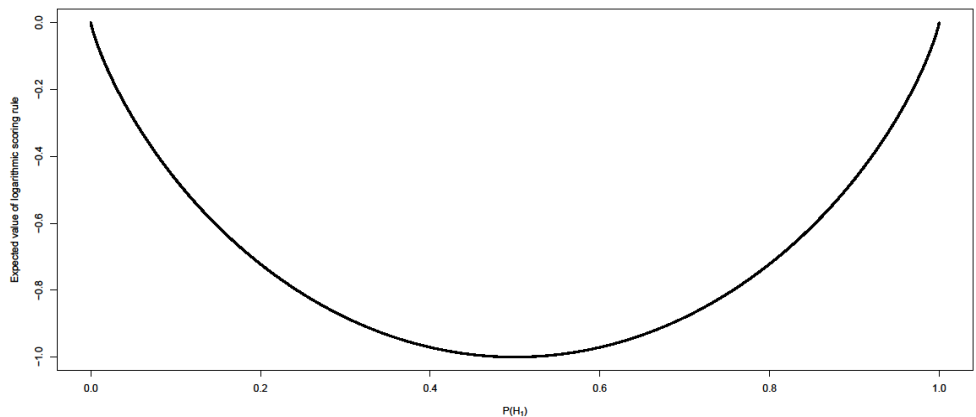

Figure 4. Expected value of logarithmic scoring rule as a function of $P(H_1)$.

Therefore, the logarithmic scoring rule adheres to the honesty and the discrimination criterion, which makes it a strictly proper scoring rule.

# Appendix 3: The law of total expectation for three random variables

The law of total expectation for two random variables $X$ and $Y$ reads:

$$B = C, \qquad \text{A3.1}$$

where $B = E_X(g(X)) = \int f(x)g(x)dx,$      A3.2

and $C = E_Y\big(E_X(g(X))\big) = \int\int f(y)f(x|y)g(x)dxdy,$      A3.3

with densities $f$ and a function $g$ on the $X$-domain.

See Gelman et al., 2003, p.23, for a proof.

Now we use this on the random variables $X = (H_A, \Delta)$ and $Y$, and focus on the situation where $g(X)$ is a strictly proper scoring rule:

$$g(X) = g(H_A, \Delta) = SPSR\big(P(H_A|\Delta)\big),$$

for example the logarithmic strictly proper scoring rule: $g(H_A, \Delta) = \log_2\big(P(H_A|\Delta)\big)$

Substituting this into Eq. 3.2, we get

$$B = \int f(H_a, \delta) SPSR\big(P(H_a|\delta)\big) d(H_a, \delta), \qquad \text{A3.4}$$

And since $A$ is countable and finite, we can rewrite $f(H_A, \Delta) = P(H_A|\Delta)f(\Delta)$, yielding:

$$A3.4 = \int \sum_a P(H_a|\delta) f(\delta) SPSR\big(P(H_a|\delta)\big) d\delta, \qquad \text{A3.5}$$

Similarly, we see that Eq. 3.3 turns into:

$$\int f(y) \int \sum_a P(H_a|\delta, y) f(\delta|y) SPSR\big(P(H_a|\delta)\big) d\delta dy, \qquad \text{A3.6}$$

According to A3.1, A3.5 = A3.6.

Q.E.D.

Ferrer, L., McLaren, M., Brummer, N., 2021. A Speaker Verification Backend with Robust Performance across Conditions. arXiv:2102.01760 [cs].

Franco-Pedroso, J., Ramos, D., Gonzalez-Rodriguez, J., 2016. Gaussian mixture models of between-source variation for likelihood ratio computation from multivariate data. PLOS ONE 11, e0149958. https://doi.org/10.1371/journal.pone.0149958

Garton, N.M., 2020. Score-based likelihood ratios and sparse Gaussian processes. PhD. Thesis. Iowa State University. https://doi.org/10.31274/etd-20200624-117

Gelman, A., Carlin, J.B., Stern, H.S., Rubin, D.B., 2003. Bayesian Data Analysis, Second Edition. CRC Press.

Geoffrey, S.M., Ewald, E., Ramos, D., González-Rodríguez, J., Lozano-Díez, A., 2020. Statistical Models in Forensic Voice Comparison, in: Banks, D., Kafadar, K., Kaye, D.H., Tackett, M. (Eds.), Handbook of Forensic Statistics. Chapman and Hall/CRC, pp. 451–497. https://doi.org/10.1201/9780367527709-20

Gneiting, T., Raftery, A.E., 2007. Strictly proper scoring rules, prediction, and estimation. Journal of the American statistical Association 102, 359–378.

Hendricks, J., Neumann, C., Saunders, C., 2018. Quantifying the weight of fingerprint evidence using an ROC-based Approximate Bayesian Computation algorithm.

Hepler, A.B., Saunders, C.P., Davis, L.J., Buscaglia, J., 2012. Score-based likelihood ratios for handwriting evidence. Forensic Science International 219, 129–140. https://doi.org/10.1016/j.forsciint.2011.12.009

Iyer, H.K., Lund, S.P., 2017. Likelihood ratio as weight of forensic evidence: A closer look. https://arxiv.org/pdf/1704.08275.pdf (accessed 27.03.23).

Leegwater, A.J., Meuwly, D., Sjerps, M., Vergeer, P., Alberink, I., 2017. Performance study of a score-based likelihood ratio system for forensic fingermark comparison. Journal of Forensic Sciences 62, 626–640. https://doi.org/10.1111/1556-4029.13339

Levey, C., 2019. OSAC Standards Bulletin, April 2019 [WWW Document]. NIST. URL https://www.nist.gov/news-events/news/2019/04/osac-standards-bulletin-april-2019 (accessed 12.23.20).

Lindley, D.V., 1991. Making Decisions. Wiley.

Lindley, D.V., 1977. A Problem in Forensic Science. Biometrika 64, 207. https://doi.org/10.2307/2335686

McElreath, R., 2020. Ch2. Small worlds and large worlds, in: Statistical Rethinking. Chapman & Hall.

Meester, R., Slooten, K., 2020. An epistemic interpretation of the posterior likelihood ratio distribution. Law, Probability and Risk 19, 139–155. https://doi.org/10.1093/lpr/mgaa010

Morrison, G.S., Enzinger, E., 2018. Score based procedures for the calculation of forensic likelihood ratios – Scores should take account of both similarity and typicality. Science & Justice 58, 47–58. https://doi.org/10.1016/j.scijus.2017.06.005

Morrison, G.S., Enzinger, E., Hughes, V., Jessen, M., Meuwly, D., Neumann, C., Planting, S., Thompson, W.C., van der Vloed, D., Ypma, R.J.F., Zhang, C., Anonymous, A., Anonymous, B., 2021. Consensus on validation of forensic voice comparison. Science & Justice 61, 299–309. https://doi.org/10.1016/j.scijus.2021.02.002

Morrison, G.S., Stoel, R.D., 2014. Forensic strength of evidence statements should preferably be likelihood ratios calculated using relevant data, quantitative measurements, and statistical models – a response to Lennard (2013) Fingerprint identification: how far have we come? Australian Journal of Forensic Sciences 46, 282–292. https://doi.org/10.1080/00450618.2013.833648

Neumann, C., Ausdemore, M., 2020. Defence against the modern arts: the curse of statistics—Part II: 'Score-based likelihood ratios.' Law Probability and Risk 19, 21–42. https://doi.org/10.1093/lpr/mgaa006

Neumann, C., Champod, C., Puch-Solis, R., Egli, N., Anthonioz, A., Meuwly, D., Bromage-Griffiths, A., 2006. Computation of likelihood ratios in fingerprint identification for configurations of